\documentclass[12pt]{article}
\input psfig.sty 
\begin{document}
\def\slash#1{\setbox0=\hbox{$#1$}#1\hskip-\wd0\hbox to\wd0{\hss\sl/\/\hss}}
\def\bea{\begin{eqnarray}}
\def\eea{\end{eqnarray}}

\begin{center}{\bf \large Two Graviton Production at $e^+e^-$ and Hadron 
Hadron Colliders\\ in the Randall-Sundrum Model}
\end{center}

\bigskip
\centerline{\bf\large Pankaj Jain and Sukanta Panda}

\begin{center} Physics Department\\
IIT, Kanpur, India 208016
\end{center}

\bigskip
\noindent
{\bf Abstract}
We compute the pair production cross section of two Kaluza Klein 
modes in the Randall-Sundrum model 
at $e^+e^-$ and hadron hadron colliders. These processes are
interesting because they get dominant
contribution from the graviton interaction at next to leading
order. Hence they provide a nontrivial test of the low
scale gravity models. All the Feynman rules at next to leading
order are also presented. These rules may be useful for many
phenomenological applications including the computation of higher
order loop corrections. 

\section{Introduction}
The proposed existence of large extra dimensions  \cite{add,add1}  
might provide a solution to 
the hierarchy problem since it can considerably lower the 
scale of quantum gravity. 
In this model the standard model fields are assumed to be confined on a
$3+1$ dimensional hypersurface or a brane in a higher dimensional space-time
\cite{Akama}. 
In this low scale gravity model, which is generally refered to as 
the ADD model, 
gravity may become strong even at TeV scale
which is far below the Planck scale $10^{19}$ GeV. The two scales 
are related via gauss law given by $M_{pl}^2$ = $M_D^{2+n} V_n$, where
$V_n$ is the volume of n extra dimensional space. If we take $M_D \sim$ TeV
then the size of the extra dimension lies in the range 1 mm - 10 fm for $n$ =
2 - 6. However by doing so we introduce a new hierarchy between the
compactification scale and the fundamental scale in the theory. An alternative
low scale gravity model (the RS model) \cite{rs} involves the existence of a
single extra dimension in a warped geometry.
In this case the scale of quantum gravity can be as small as 1 TeV
without requiring the existence of very large extra dimensions. 
These low scale gravity models are reviewed in Ref. \cite{Kubyshin,Hewett}.
The leading order Feynman rules for the ADD model are given 
in Lykken et al \cite{hlz}
and Giudice et al \cite{grw}.

Phenomenologically the  RS model \cite{dhr} is very different from ADD model 
because of the presence of very massive KK modes which
can produce observable 
signatures in future collider experiments.  
In the case of ADD model we instead have large number of 
closely spaced KK modes
with the mass the lightest mode governed by the size of the large extra
dimension. These modes cannot be observed directly due to their 
small coupling to matter and hence the production of these modes gives rise to
missing energy signatures. However in the RS model the
first few modes  produced can decay into observable particles.
The center of mass energy required to produce these resonances are
well within the energies available in the next generation of
colliders such as the LHC. Non-observation 
of these processes in experiments constrain the parameters of  
this theory. Several phenomenological studies have been carried
out to study these resonances in $e^+e^-$ colliders 
\cite{Rizzo,Das,Ghosh,rai03} and in hadron colliders 
\cite{Allanach,Bijnens,Allanach1,Traczyk,Dvergsnes}.
 
In the present paper we study the process of two graviton production
in the RS model in $e^+ e^-$ and hadron hadron colliders. This
process gets contribution at next to leading order in the gravitational
coupling. Hence we derive the required Feynman rules at this order. 
This process may 
not be very important in case of ADD model because these modes  just
escape through the detector undetected. However our calculation
can be easily generalized for this case also. 

The plan of the rest of the paper is as follows.
In the
second section we briefly review the derivation of lowest order interaction
Lagrangian. We then extend the
calculation to next order in coupling and find the 
corresponding interaction Lagrangian.  
Third section describes all processes related to emission of two lowest
massive graviton modes. Here we also present our results for
the cross section of two graviton emission at $e^+ e^-$ and hadron hadron
colliders. Section four gives our conclusions.  
In Appendix A we present all the Feynman rules involving vertex 
of matter fields with
two gravitons at next to leading order in
coupling.         
In Appendix B, we derive the Feynman rule for three graviton vertex which
also contributes to the process under consideration.  

\section{Interaction Lagrangian}
The RS model assumes a five dimensional non-factorizable geometry
where two $3-branes$, with opposite tensions, reside at $S^1/Z_2$ orbifold
fixed points. In this
framework solution to  $5$-dimensional Einstein equations is given by
 
\vspace*{0.05in}

\begin{equation}
ds^2 = e^{-\sigma(\phi)}\eta_{\mu\nu}dx^{\mu}dx^{\nu}+r_{c}^2 d\phi^2
\end{equation}
where $\phi$ is the angular coordinate parameterizing the extra
dimension, $\sigma(\phi)=kr_c|\phi|$ with $r_c$ being the compactification
radius of the extra dimension, and $0\leq|\phi|\leq\pi$. Starting from the
5-D action, after integrating over extra coordinate $\phi$, we obtain the
reduced effective 4-D planck scale given by the relation
\begin{equation}
M_{Pl}^2 = \frac{M^3}{k}(1-e^{-2kr_c\pi})\ .
\label{massscale}
\end{equation}
 Setting visible brane at $\phi=\pi$, where our standard
model fields live, it is found that any mass parameter $\tilde m$ on the
visible
3-brane in the fundamental higher-dimensional theory will correspond to a
physical mass $$m=e^{- k r_c\pi}\tilde m\ .$$
Hence TeV mass scales can be generated on
the 3-brane at $\phi=\pi$ due to exponential factor present in the metric
if we assume $kr_c \approx 12$. 

A small perturbation $h_{\mu\nu}$ of the metric, 
such that 
$ e^{-2\sigma(\phi)}\eta_{\mu\nu} \rightarrow  e^{-2\sigma(\phi)}(\eta_{\mu\nu}
+h_{\mu\nu})$, can be expanded as, 
\begin{equation}
h_{\mu\nu}(x,\phi)=\sum_{n=0}^{\infty}h_{\mu\nu}^{(n)}\frac{\chi^{(n)}(\phi)}
{\sqrt r_c}\ . 
\end{equation}
After solving the linearised equation of motion
of graviton field and working in
the gauge  $$\partial^{\mu}h_{\mu\rho}^{(n)}=h_{\mu}^{\mu(n)} = 0, $$ one
obtains the solution for $\chi^{(n)}(\phi)$, 
\begin{equation}
\chi^{(n)}(\phi) =
\frac{e^{2\sigma(\phi)}}{N_n}J_2\left(\frac{M_n}{k}
e^{\sigma(\phi)}\right)+\alpha_nY_2\left(\frac{M_n}{k}
e^{\sigma(\phi)}\right)
\ .
\end{equation}
The exact expression for
$\alpha_n$ is given in
Ref. \cite{dhr}. The masses of the KK modes, $M_n$, are given by 
\cite{dhr,Goldberger}
\begin{equation}
M_n = x_n k e^{- kr_c\pi} \,
\label{mode}
\end{equation}
where $x_n$s are the solution of the equation $J_1(x_n) = 0.$
We also define the mass parameter $m_0= M_n/x_n$. 
As we see from Eq. \ref{mode}, masses of the graviton KK excitations are
dependent on the roots of $J_1$ and are not equally spaced. 
In the following, we formulate the interaction 
of physical KK modes to the Standard Model fields which
reside on the brane at $\phi=\pi$.

To start with, let us first consider  the minimal gravitational coupling
of the general scalar S, vector V and fermion F,
\begin{equation}
\int d^4 x  \sqrt{-\hat{g}} \mathcal{L}(\hat{g},S,V,F)
\end{equation}
where the induced metric $\hat{g}$ on the brane can be decomposed as, 
$$\hat g_{\mu \nu}  =  \eta_{\mu \nu} + \kappa h_{\mu \nu}\ .$$ 
By expanding the interaction Lagrangian upto ${\mathcal{O}}(\kappa^{2})$,
we get
\begin{eqnarray}
\int d^4 x \int d\phi  \sqrt{-\hat{g}} {\mathcal{L}}(\hat{g})
&=&
\int d^4 x \int d\phi \delta(\phi-\pi)\left[ 
{\mathcal{L}}(\hat{g})|_{\hat{g} =
\eta} - \frac
{\kappa}{2} 
 h^{\mu\nu}T_{\mu\nu}  
\right]
\nonumber \\ 
& &
+ \kappa^2  \left[ A
{\mathcal{L}}(\hat{g})|_{\hat{g}
= \eta} 
- B^{{\mu}{\nu}}
\frac{\delta{\mathcal{L}}}{\delta\hat{g}^{\mu\nu}}|_{\hat{g} 
= \eta} \right]
\nonumber \\
& & \left.
+ \kappa^2  h^{\mu\nu}(x,\phi)\left[
\frac{1}{2}\int d^4 y
\frac{\delta^2
{\mathcal{L}}}
{{\delta\hat{g}^{\mu\nu}}{\delta\hat{g}^{\rho\sigma}}}|_{\hat{g} 
= \eta} h^{\rho\sigma}(y,\phi)\right]\right]
\nonumber \\
& &
\label{lan}
\end{eqnarray}
where we have used 
\begin{eqnarray}
\hat g^{\mu \nu}  =  \eta^{\mu \nu} - \kappa h^{\mu \nu} 
 + \kappa^2 h^{\mu \rho} h_{\rho}^{\nu}  
 \\
\sqrt{-\hat {g}}  =  1 + \frac{\kappa}{2}h  
+  \kappa^2(\frac{1}{8}h^2
 - \frac{1}{4}h_{\rho \sigma}h^{\rho \sigma}) 
\end{eqnarray}
Several coefficients that appears in  Eq. (\ref{lan}) are given by,
\begin{eqnarray}
A = \frac{1}{8}h^2  
- \frac{1}{4}h_{\rho\sigma}h^{\rho\sigma}\ , \\
B^{\mu\nu} = \frac{1}{2}h h^{\mu\nu} - h^{\mu\lambda}h_{\lambda}^{\nu}\ , \\
h = h^{\mu}_{\mu} = h^{\mu\nu} \eta_{\mu\nu}\ .
\end{eqnarray}
Interaction Lagrangian upto ${\mathcal{O}}(\kappa)$ is given by
$$-\frac{\kappa}{2}h^{\mu\nu}(x,\phi=\pi)T_{\mu\nu}=-\frac{1}{M^{3/2}}
h^{\mu\nu}(x,\phi=\pi)T_{\mu\nu},$$ 
where $\kappa = \frac{2}{M^{3/2}}$. With the help of Eq. \ref{massscale}
we can express the above interaction Lagrangian in terms of KK modes as,
\begin{equation}
-\frac{1}{M_{Pl}}h^{\mu\nu,0}T_{\mu\nu}-\frac{1}{\Lambda_{\pi}}
\sum_{n=1}^{\infty}h^{\mu\nu,n}T_{\mu\nu}\,
\end{equation}
where $T_{\mu\nu}$, the energy-momentum tensor of the matter field
confined on the brane, is given by,
\begin{equation}
T_{\mu\nu} = - \eta_{\mu\nu} {\mathcal{L}}(\hat g)|_{\hat g = \eta} + 2
\frac
{\delta{\mathcal{L}}}{\delta \hat g^{\mu\nu}}|_{\hat g = \eta}
\end{equation}
and $$\Lambda_{\pi} = M_{Pl} e^{-kr_c\pi}\ .$$
Now we are interested in finding out the ${\mathcal{O}}(\kappa^2)$
Feynman rules for the gravitational interaction with matter confined
on the $D_3$ brane embedded in the RS  bulk. 

The $\kappa^2$ part of the Lagrangian
that resembles the gravitational interactions with matter involving two
gravitons, is given by
\begin{eqnarray}
{\mathcal{L}}_2 &  = & \frac{4}{M^3}\left[
\sum_{n=0}^{\infty}\sum_{m=0}^{\infty}
h^{\mu
\lambda,n}\chi^n(\phi=\pi)
h_{\lambda}^{\nu,m}\chi^m(\phi=\pi) 
\left. \frac {\delta 
{\mathcal{L}}} {\delta\hat g^{\mu\nu}}\right|_{\hat g = 
\eta}
\right.
\nonumber \\
& &
-(1/2)\sum_{n=0}^{\infty}\sum_{m=0}^{\infty}  h^{n}\chi^n(\phi=\pi) h^{\mu
\nu,m}\chi^m(\phi=\pi)\left. 
\frac {\delta{\mathcal{L}}} {\delta\hat g^{\mu\nu}}\right|_{\hat g = \eta}
\nonumber \\
& & 
+ (1/8)\sum_{n=0}^{\infty}\sum_{m=0}^{\infty}  h^{n}\chi^n(\phi=\pi)
h^{m}\chi^m(\phi=\pi)
\left.{\mathcal{L}}(\hat g)\right|_{\hat g = \eta} 
\nonumber \\
& &
-  (1/4)\sum_{n=0}^{\infty}\sum_{m=0}^{\infty}  h_{\rho
\sigma}^n\chi^n(\phi=\pi) h^{\rho
\sigma,m}\chi^m(\phi=\pi)\left.
{\mathcal{L}}(\hat g)\right|_{\hat g = \eta}
\nonumber \\
& & \left. 
+ (1/2)\sum_{n=0}^{\infty}\sum_{m=0}^{\infty}  h^{\mu
\nu,n}\chi^n(\phi=\pi) \left. \frac {\delta^{2}
{\mathcal{L}}} {{\delta\hat g^{\mu\nu}}  {\delta\hat g^{\rho \sigma}}}
\right|_{\hat g = \eta}  h^{\rho \sigma,m}\chi^m(\phi=\pi)\right]\ .
\end{eqnarray}
After substituting the expression of $\chi^n$  in
the above equation and using the relation given in Eq. \ref{massscale}, we
finally  get
\begin{equation}
{\mathcal{L}}_2 = {\mathcal{L}}_2^{00} +
2 {\mathcal{L}_2}^{0n}+{\mathcal{L}}_2^{nm}
\end{equation}
where the ${\mathcal{L}}_2^{nm}$ part of the lagrangian is given by

\begin{eqnarray}
{\mathcal{L}}_2^{nm} &  = & \frac{4}{\Lambda_\pi^2}
\sum_{n=1}^{\infty}\sum_{m=1}^{\infty}\left[
h^{\mu
\lambda,n}
h_{\lambda}^{\nu,m} 
\left. \frac {\delta 
{\mathcal{L}}} {\delta\hat g^{\mu\nu}}\right|_{\hat g = 
\eta}
\right.
\nonumber \\
& &
-(1/2)  h^{n} h^{\mu
\nu,m}\left. 
\frac {\delta{\mathcal{L}}} {\delta\hat g^{\mu\nu}}\right|_{\hat g = \eta}
\nonumber \\
& & 
+ (1/8)  h^{n}
h^{m}
\left.{\mathcal{L}}(\hat g)\right|_{\hat g = \eta} 
\nonumber \\
& &
-  (1/4)  h_{\rho
\sigma}^n h^{\rho
\sigma,m}\left.
{\mathcal{L}}(\hat g)\right|_{\hat g = \eta}
\nonumber \\
& & \left. 
+ (1/2)  h^{\mu
\nu,n} \left. \frac {\delta^{2}
{\mathcal{L}}} {{\delta\hat g^{\mu\nu}}  {\delta\hat g^{\rho \sigma}}}
\right|_{\hat g = \eta}  h^{\rho \sigma,m}\right]\ .
\label{la}
\end{eqnarray}
The ${\mathcal{L}}_2^{00}$ part of the lagrangian is obtained by setting 
$n=m=0$ and by replacing the coefficient $\Lambda_\pi^2$ by $M_{Pl}^2$ 
in ${\mathcal{L}}_2^{nm}$. Due to the large suppression factor 
$1/M_{Pl}^2$ the contribution of this lagrangian as well as that of 
$ {\mathcal{L}_2}^{0n}$ is negligible. 
By using the interaction Lagrangian, we can derive the feynman rules
for two gravitons interacting with matter fields which are given in
Appendix A of this paper. Also the feynman rules for three graviton vertex
are presented in Appendix B. The triple graviton vertex has also
been studied in Ref. \cite{dr}. 
\begin{figure}[tb]
\hbox{\hspace{0em}
\hbox{\psfig{figure=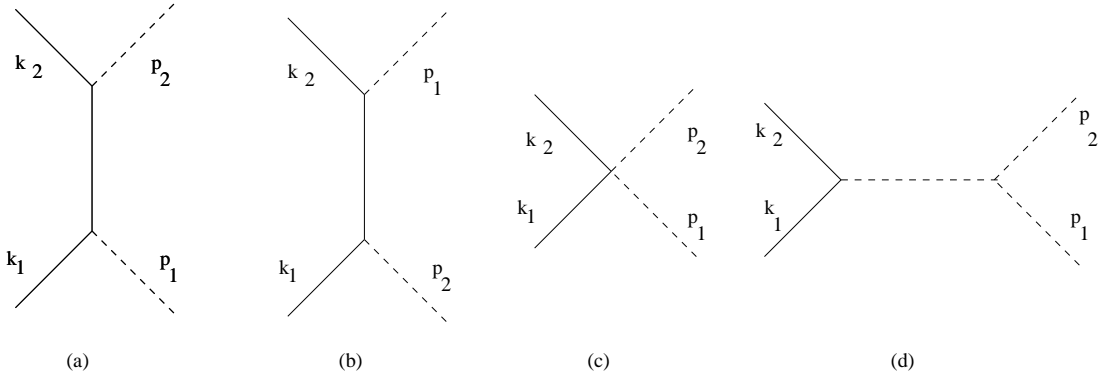,height=6cm}}}
\caption {All possible tree level Feynman diagrams
which contribute to two graviton emission due to fermion 
anti-fermion annihilation. The
solid and dotted lines represent the fermions and the  gravitons
respectively.} 
\label{fig:diagrams}
\end{figure}                                                                                        
\section{Two Graviton Production}
\begin{figure}[tb]
\hbox{\hspace{0em}
\hbox{\psfig{figure=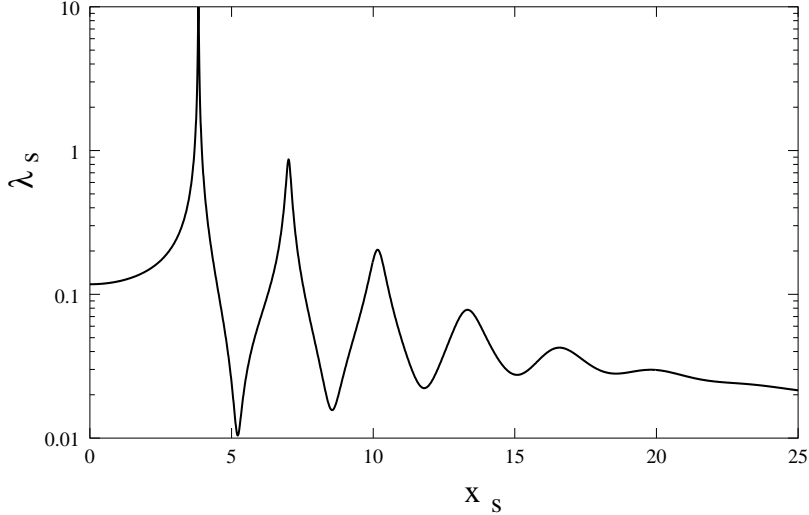,height=8cm}}}
\caption { The graviton propagator factor $\lambda_s$ as a function of 
$x_s=\sqrt{s}/m_0$. Here we have fixed the
parameters $m_0=100$ GeV and $c_0=.01$.}
\label{fig:lambdas}
\end{figure}

We next consider the production of two gravitons at $e^+ e^-$,
$p\bar p$ and $pp$ colliders. We first consider the process
fermion anti-fermion annihilation into two gravitons. 
There are four feynman diagrams contributing to this  
which are shown in
Fig. (\ref{fig:diagrams}). 
The differential cross section for $e^+e^-\rightarrow GG$ is given in 
Eq. \ref{Eq:ffbartoGG} in Appendix C. 
The cross section also gets contribution from the s-channel diagram
which involves a graviton propagator. This gets contribution from all
the Kaluza-Klein modes. By summing over all the contributions 
 the propagator takes the
form
\begin{equation}
D(s) = \frac{32 \pi c_0^2}{m_0^2} \sum_{n}\frac{1}{s- M_n^2 + i \Gamma_n
M_n} =\frac{32 \pi c_0^2}{m_0^4} \lambda_s(x_s).
\end{equation}
where $x_{s}=\frac{\sqrt{s}}{m_0}$ and $c_0 = k/M_{\rm pl}$. 
By taking into account all decay
channels, $\Gamma_n$ can be written as
\begin{equation}
\Gamma_n = c_0^2 m_0 x_n^3 \Delta(M_n)
\end{equation}
Here $\Delta$ is a complicated function of $M_n$. 
We calculate the $\lambda_s(x_s)$
numerically and it's behaviour is shown in Fig {\ref{fig:lambdas}}.

\begin{figure}[tb]
\hbox{\hspace{0em}
\hbox{\psfig{figure=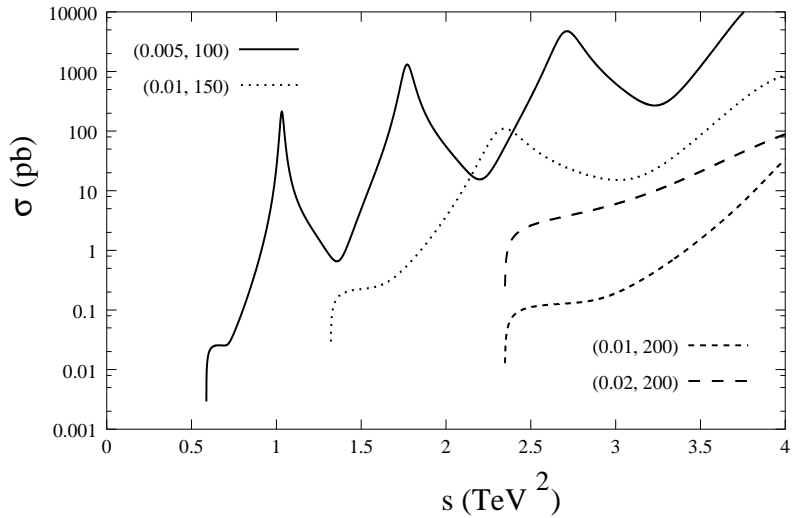,height=8cm}}}
\caption {The two graviton production cross section 
in $e^+e^-$ collisions as a function of
the center of mass energy squared $s$ (TeV$^2$) for several
different values of the parameters $(c_0,m_0)$, where $m_0$ is given
in GeV. 
}
\label{fig:ees}
\end{figure}

\begin{figure}[tb]
\hbox{\hspace{0em}
\hbox{\psfig{figure=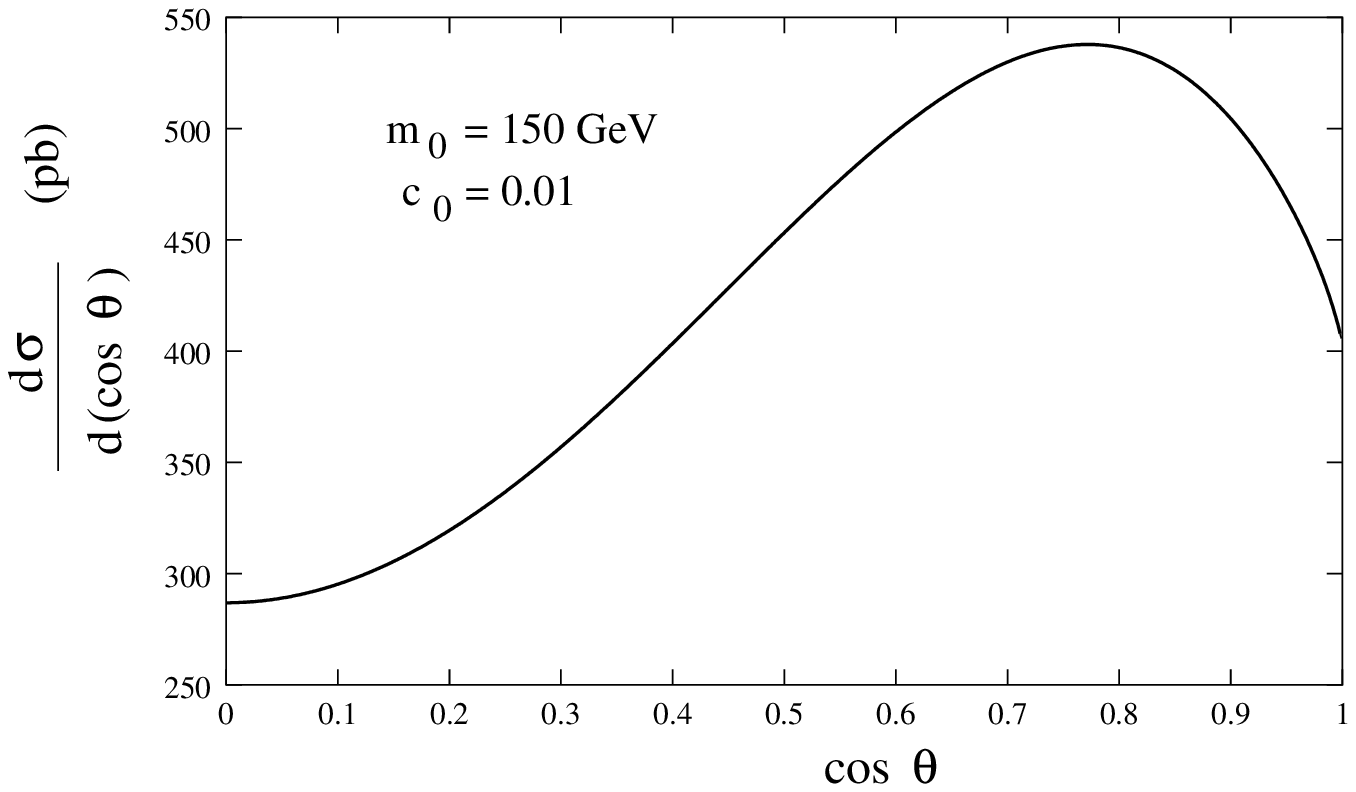,height=8cm}}}
\caption {The differential cross section for $\sqrt{s}=2$
TeV as a function of $\cos\theta$ for the production of the first KK
mode in $e^+ e^-$ collisions. Here $\theta$ is the center of mass
scattering angle and
we have fixed $c_0=0.01$, $m_0=150$ GeV.}
\label{fig:eecostheta}
\end{figure}

In Fig. \ref{fig:ees} we show the $s$ dependence of the 
production cross section of the two lightest KK modes of mass $M_1 = 3.83 m_0$ 
in $e^+e^-$ collisions. 
The resonance peaks which arise due to the s-channel graviton exchange
(diagram \ref{fig:diagrams}d) are clearly visible in a certain range of
parameter space. The peaks tend to disappear for large values of the
parameter $c_0=k/M_{\rm pl}$. We expect that the first few peaks
will be identifiable if the parameter $c_0$ is sufficiently small. 
For values of $c_0$ larger than approximately 0.6 even the first
resonance may not be clearly identifiable \cite{rai03}. 
The maximum value of the parameter $m_0$ that can be explored with 
this process is given by $\sqrt{s} <2M_1=7.66 m_0$. 
In Fig. {\ref{fig:eecostheta}}, we show the angular dependence of 
differential cross section for pair production of the first graviton
mode with mass $M_1=3.83m_0$ where $m_0$ has been chosen to be 150 GeV.

We see from fig. \ref{fig:ees} that, as expected, 
the cross section rises very rapidly
with $s$. The perturbation theory is applicable
only if $s<<\Lambda_\pi^2$. In our calculations we have chosen
the parameter space such that this condition is respected and 
perturbative predictions can be trusted. Perturbation theory breaks
down for values of $c_0$ much larger than
those chosen in fig. \ref{fig:ees} for the corresponding values of
$m_0$. In this region it is not possible to compute this process reliably
with our current understanding of quantum gravity.  

\begin{figure}[tb]
\hbox{\hspace{0em}
\hbox{\psfig{figure=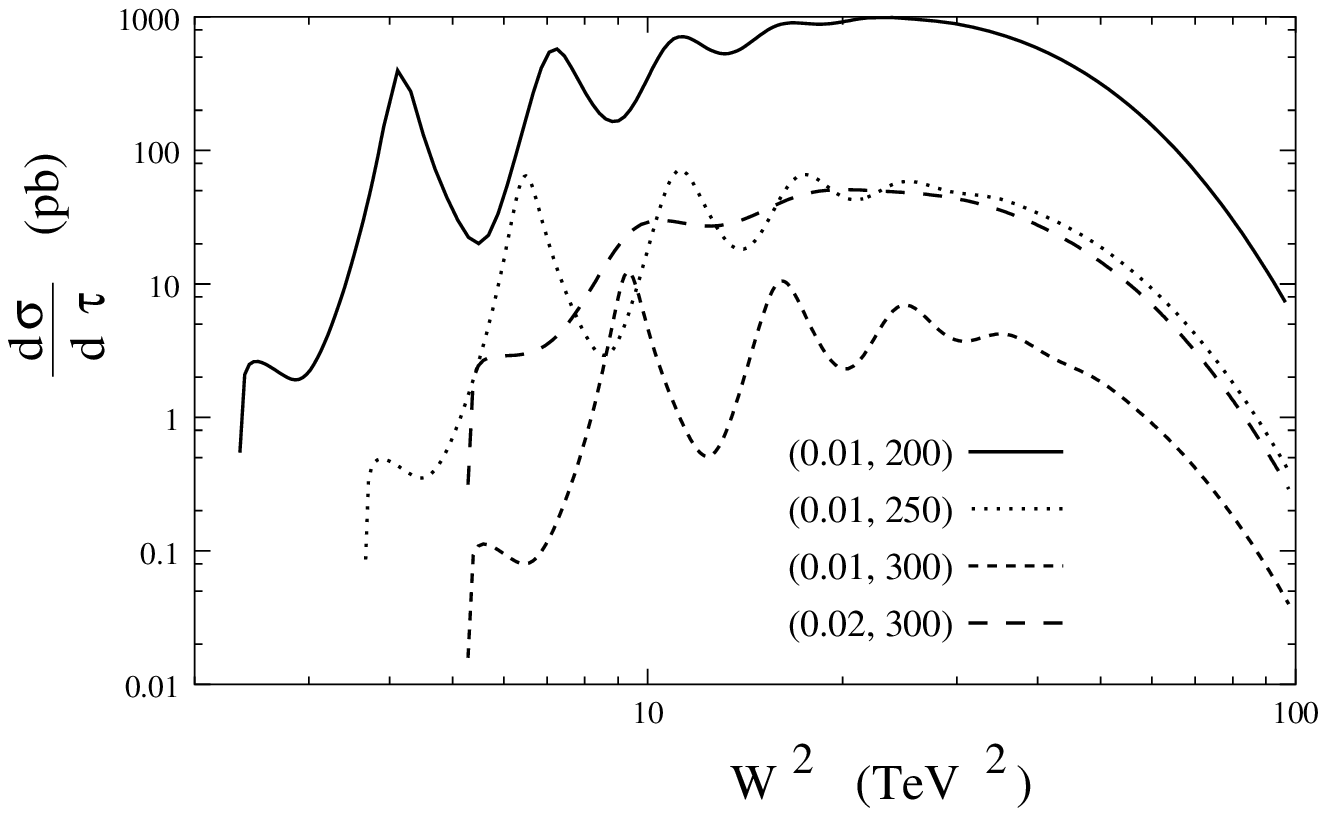,height=8cm}}}
\caption {The invariant mass $W^2$ dependence of the two graviton 
production cross section in proton proton collisions at $\sqrt {s} = 14$ 
TeV for several
different values of the parameters $(c_0,m_0)$, where $m_0$ is given
in GeV. Here
$W^2=\tau s$ is the invariant mass of the two graviton final state. }
\label{fig:pp_W2}
\end{figure}

\begin{figure}[tb]
\hbox{\hspace{0em}
\hbox{\psfig{figure=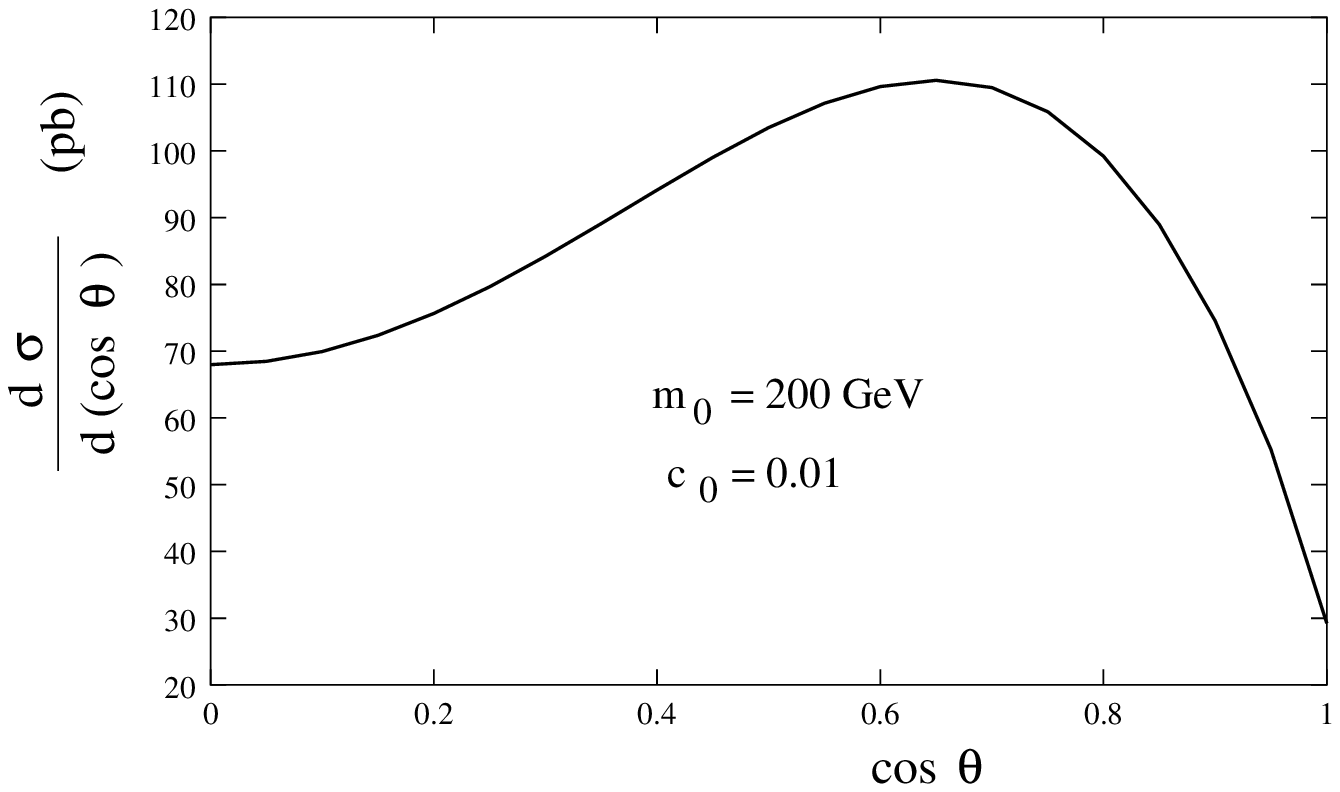,height=8cm}}}
\caption {The $\cos\theta$ dependence of the two graviton 
production cross section in proton proton collisions at $\sqrt {s} = 14$
TeV. Here
$\theta$ is the center of mass scattering angle of the partonic 
subsystem.}
\label{fig:pp_costh}
\end{figure}

\begin{figure}[tb]
\hbox{\hspace{0em}
\hbox{\psfig{figure=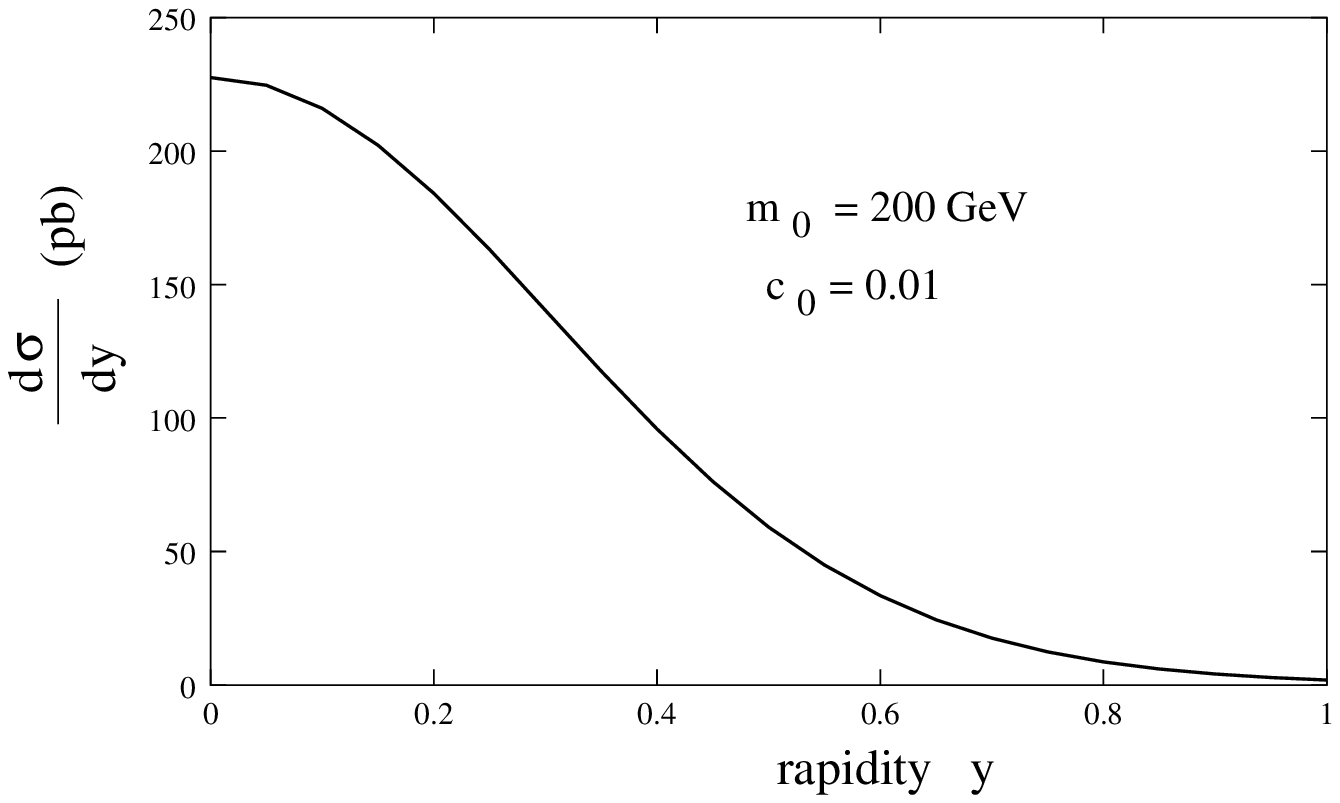,height=8cm}}}
\caption {The rapidity dependence of the two graviton  
production cross section in proton proton collisions at $\sqrt {s} = 14$
TeV. 
}
\label{fig:pp_rapidity}
\end{figure}

\begin{figure}[tb]
\hbox{\hspace{0em}
\hbox{\psfig{figure=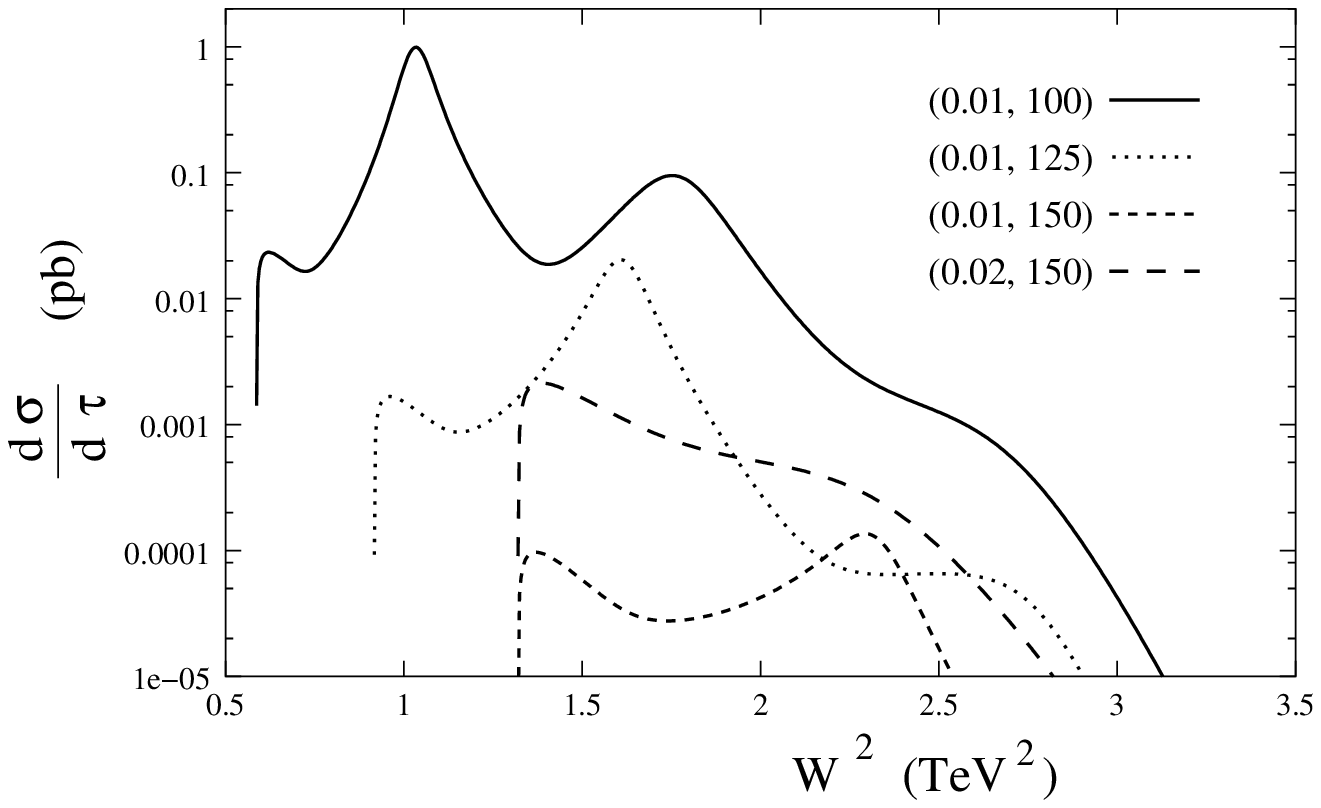,height=8cm}}}
\caption {The invariant mass $W^2$ dependence of the two graviton 
production cross section in proton anti-proton collisions at $\sqrt {s} = 2$ 
TeV for several
different values of the parameters $(c_0,m_0)$, where $m_0$ is given
in GeV. Here
$W^2=\tau s$ is the invariant mass of the two graviton final state. }
\label{fig:ppbar_W2}
\end{figure}

\begin{figure}[tb]
\hbox{\hspace{0em}
\hbox{\psfig{figure=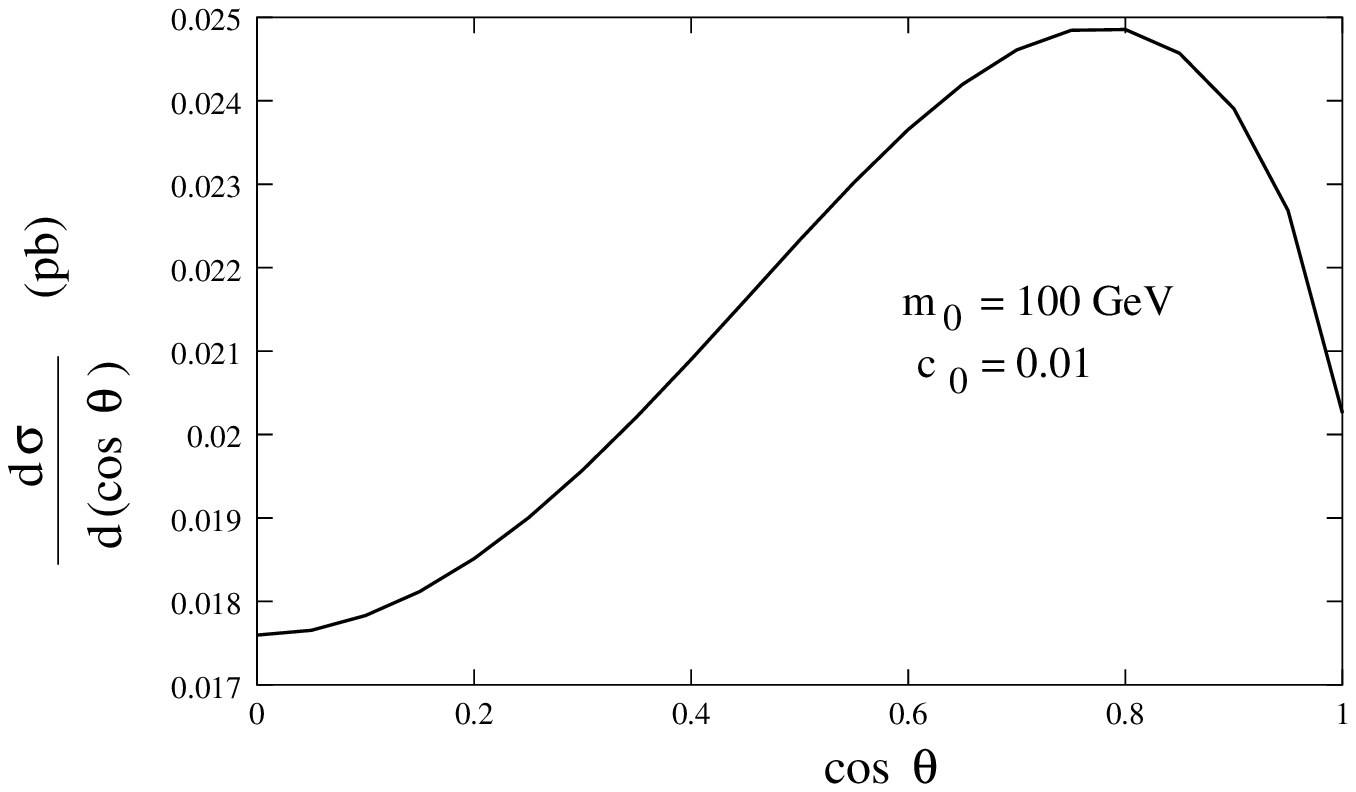,height=8cm}}}
\caption {The $\cos\theta$ dependence of the two graviton 
production cross section in proton anti-proton collisions at $\sqrt {s} = 2$
TeV. Here
$\theta$ is the center of mass scattering angle in the partonic subsystem.}

\label{fig:ppbar_costh}
\end{figure}

\begin{figure}[tb]
\hbox{\hspace{0em}
\hbox{\psfig{figure=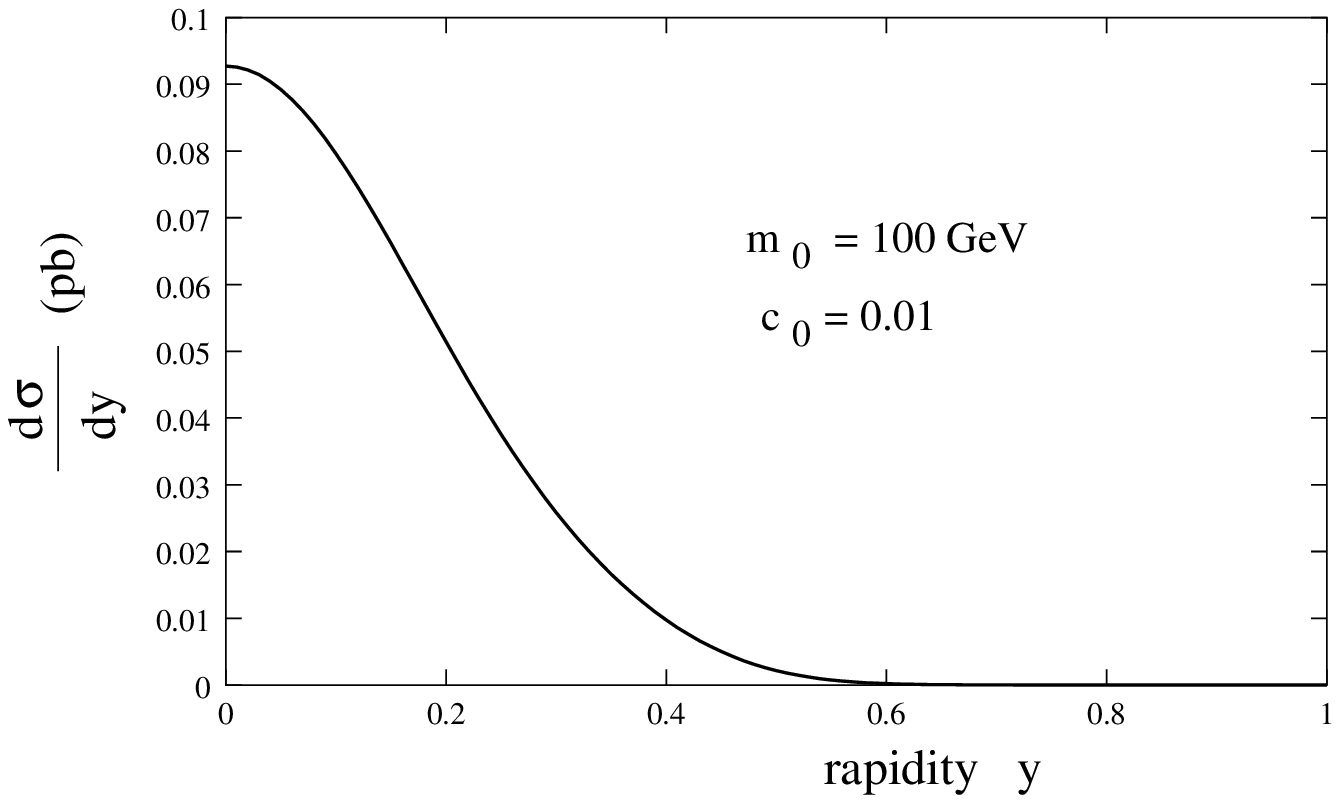,height=8cm}}}
\caption {The rapidity dependence of the two graviton  
production cross section in proton anti-proton collisions at $\sqrt {s} = 2$
TeV.  }
\label{fig:ppbar_rapidity}
\end{figure}

The hadron-hadron crosss section gets contributions both from quarks
and gluons. In the case of quark anti-quark annihilation the contributing
diagrams are same as listed in Fig. \ref{fig:diagrams}.
The diagrams contributing to gluon gluon fusion are also the same
as those given in Fig. \ref{fig:diagrams} with the quark lines replaced
by gluons. The results for the pair production of the lightest KK mode
with mass $M_1 = 3.83 m_0$ is given in figures \ref{fig:pp_W2}, 
\ref{fig:pp_costh} and \ref{fig:pp_rapidity}. 
The center of mass energy $\sqrt s$ of the hadron hadron
system is taken to be the LHC energy, i.e. 14 TeV.
In fig. \ref{fig:pp_W2}
we show the $W^2$ dependence of the cross section, where $W^2=\tau s=x_1x_2s$ is
the invariant mass of the final state and $x_1$ and $x_2$ are the longitude
momentum fractions of the two initial state partons. 
We see from this figure that, as expected, LHC can probe a 
considerable range of parameters of the RS model through this process. 
In figs. 
\ref{fig:pp_costh} and \ref{fig:pp_rapidity} we show the 
$\cos\theta$ and the rapidity $y={1\over 2}\log(x_1/x_2)$
dependence of the cross section respectively. Here $\theta$ is
the center of mass scattering angle 
in the parton subsystem.
In Figs. \ref{fig:ppbar_W2},
\ref{fig:ppbar_costh} and \ref{fig:ppbar_rapidity} we show the corresponding
results for the proton anti-proton system at center of mass energy
$\sqrt s=2$ TeV, relevant for the Tevatron.

Each of the two gravitons produced in the final state eventually decay into
either two jets, a lepton pair or a vector boson pair. Therefore one possible  
experimental signature of this process involves a four jet event
such that the invariant mass of the two pairs of jets is equal to
the mass of the graviton. There will be considerable QCD background
which may limit the parameter space that can be explored with this process.
We postpone the detailed study of experimental signatures to future research.

\section{Conclusions}
We have computed the pair production cross section of Kaluza Klein modes
in the RS model in $e^+e^-$ and hadron hadron colliders at leading
order in the gravitational coupling $\kappa$. The leading order calculation
requires the Feynman rules at order $\kappa^2$.  
This process displays a resonance structure due to the contribution
from the s-channel KK mode exchange. 
For a large range of the parameter space this resonance structure is 
visible in the two graviton production channel. The gravitons in the final
state can be identified through their decay products which include a 
pair of jets, a lepton pair or a vector boson pair. 
We find that LHC can study a fairly large range of the parameter
space of the RS model with this process. The process provides a nontrivial
test of the RS model since its calculation requires the next to leading 
order Feynman rules. 

\bigskip
\noindent
{\bf Acknowledgements} We are very grateful to Prasanta Das and
Sreerup Raychaudhuri who collaborated with us during the initial
stages of this work and made considerable contributions. We also
thank Sreerup Raychaudhuri and Santosh Rai for help in the numerical
work.

\newpage

\vskip 10pt
{\Large\bf Appendix A: Coupling to Matter Fields}\\

{\large\bf{B.1 Coupling to Scalar Bosons}}

\vspace{.5cm}

 For a general complex scalar field $\Phi$,we have \\
\begin{eqnarray}
\label{ls}
{\mathcal{L}}_{s}(g) & = &g^{\rho \sigma} D_{\rho}\Phi^{\dagger}
D_{\sigma}\Phi
- m_{\Phi}^{2} \Phi^{\dagger} \Phi \\
\label{ls1}
\left. \frac {\delta
{\mathcal{L}}_{s}} {\delta\hat g^{\mu\nu}}\right|_{\hat g=\eta} & = &
\frac{1}{2}( D_{\mu} \Phi^{\dagger}
D_{\nu} \Phi + D_{\nu} \Phi^{\dagger}
D_{\mu} \Phi) \\
\label{ls2}
\left. \frac { \delta^{2}
{\mathcal{L}}_{s} }
{ \delta\hat g^{\mu\nu}  \delta\hat g^{\rho \sigma}}
\right|_{\hat g = \eta} & = & 0
\end{eqnarray}

where the gauge covariant derivative is defined as
\begin{equation}
D_{\mu} = \partial_{\mu} + ig A_{\mu}^{a}T^{a}
\end{equation}
with g the coupling, $  A_{\mu}^{a} $ the gauge fields and $T^{a}$ the Lie
algebra generators.

Substituting the expressions as given in Eq. (\ref{ls})~(\ref{ls1}) and
Eq. (\ref{ls2})in
the lagrangian given in Eq. (\ref{la}), we finally have the Feynman rule
for the 
vertex involving two gravitons, and two scalars as follows:

\begin{eqnarray}
 h_{\mu\nu}^{(n)}  h_{\rho\sigma}^{(m)} \Phi \Phi :
i \frac {1}{\Lambda_\pi^2}\left[ \left[ 
\eta_{\sigma\nu}(k_{1\mu}k_{2\rho} + k_{1\rho}k_{2\mu}) 
+ \eta_{\mu\sigma}(k_{1\rho}k_{2\nu} + k_{1\nu}k_{2\rho}) 
+ \eta_{\nu\rho}(k_{1\mu}k_{2\sigma} + k_{1\sigma}k_{2\mu})\right]
\right.
\nonumber \\
+  \left[ 
\eta_{\mu\rho}(k_{1\sigma}k_{2\nu}+k_{1\nu}k_{2\sigma}) 
+ \eta_{\mu\nu}(k_{1\rho} k_{2\sigma} + k_{1\sigma} k_{2\rho}) 
- \eta_{\rho\sigma}(k_{1\mu}k_{2\nu} + k_{1\nu}k_{2\mu})\right]
\nonumber \\
\left.
-  \left[ 
(\eta_{\mu\rho}\eta_{\nu\sigma} + \eta_{\nu\rho}\eta_{\mu\sigma} - 
\eta_{\mu\nu}\eta_{\rho\sigma}) (k_1.k_2 - m_{\Phi}^2) \right]\right]
\end{eqnarray}
This can be rearranged in the following simplified manner,
\begin{eqnarray}
i \frac {1}{\Lambda_\pi^2}  \left(C_{\mu\nu,\rho\sigma} m_{\Phi}^2 +
C_{\mu\nu,~\rho\sigma}|_{\lambda\eta} k_1^\lambda k_2^\eta \right) 
\end{eqnarray}
where $k_1$, $k_2$ are the four-momenta of the scalars,
$$C_{\mu\nu,\rho\sigma} = \eta_{\mu\rho}\eta_{\nu\sigma} +
\eta_{\mu\sigma}\eta_{\nu\rho} - \eta_{\mu\nu}\eta_{\rho\sigma}$$
and
$$C_{\mu\nu,\rho\sigma}|_{\lambda\eta} = \frac{1}{2}\left[ 
\eta_{\mu\lambda} C_{\rho\sigma,~\nu\eta} + 
\eta_{\sigma\lambda} C_{\mu\nu,~\rho\eta} + 
\eta_{\rho\lambda}C_{\mu\nu,~\sigma\eta} + 
\eta_{\nu\lambda}C_{\mu\eta,~\rho\sigma} - 
\eta_{\lambda\eta}C_{\mu\nu,~\rho\sigma} + 
(\lambda \leftrightarrow \eta)\right]\ .$$
Similarly, one can obtain the Feynman rules for the vertices involving
two gravitons,
two scalars and one gauge bosons(and two gauge bosons) as follows:
\begin{eqnarray} 
 h_{\mu\nu}^{(n)}  h_{\rho\sigma}^{(m)}
A^a_{\lambda} 
\Phi \Phi :
- \frac{i g}{ \Lambda_\pi^2} 
C_{\mu\nu,~\rho\sigma}|_{\lambda\eta} (k_1 + k_2)^\eta T^{a}_{MN}
\end{eqnarray} 
\begin{eqnarray}
 h_{\mu\nu}^{(n)}  h_{\rho\sigma}^{(m)}
A^a_{\eta}A^b_{\lambda}
\Phi \Phi :
- \frac{i g^2}{\Lambda_\pi^2}
C_{\mu\nu,~\rho\sigma}|_{\lambda\eta}  \{T^{a},T^{b}\}_{MN}
\end{eqnarray}


\vskip 10pt
\noindent
{\large\bf {B.2 Couplings To Fermions}}

\vspace{.5cm}

Vierbein formalism is required in order to incorporate the fermion
in the gravitation theory.
The Fermion lagrangian  is given by
\begin{equation}
{\mathcal{L}}_\psi(\hat g) = {\overline \psi} i \gamma^{\rho} {\mathcal{D}}_\rho \psi 
  - m_\psi {\overline \psi \psi}   
= \frac{1}{2} ({\overline \psi} i \gamma^{\rho} {\mathcal{D}}_\rho \psi 
  - {\mathcal{D}}_\rho{\overline \psi} i \gamma^{\rho} \psi) - m_{\psi} {\overline \psi} \psi 
\label{lf}
\end{equation}
The covariant derivative ${\mathcal{D}}_\rho \psi$ is given by,
\begin{eqnarray}
{\mathcal{D}}_\mu \psi = (D_{\mu} + \frac{1}{2} w_{\mu}^{ab}\sigma_{ab})\psi
\end{eqnarray}
The spin connection, $w_{\mu}^{ ab}$ can be written as,
\begin{eqnarray}
w_{\mu a b} = \frac{1}{2} \left (\partial_\mu e_{b \nu} - \partial_\nu e_{b \mu}\right )e_a^\nu 
- \frac{1}{2} \left (\partial_\mu e_{a \nu} - \partial_\nu e_{a \mu}\right )e_b^\nu
- \frac{1}{2} e_a^\rho e_b^\sigma \left ( \partial_\rho e_{c \sigma} - \partial_\sigma e_{c \rho} \right )
e^c_\mu
\end{eqnarray}
where $ e_{\mu}^{a} e_{\nu}^{b} \eta_{a b} = g_{\mu\nu}$,~$\gamma^{\mu} =
e^{\mu}_{a} \gamma^{a}$,~$e^{\mu}_a e_{b\mu} = \eta_{ab}$ and $\sigma_{ab}
= \frac{1}{4} [\gamma_{a},\gamma_{b}]$ . 
The conserved energy-momentum tensor of the fermion field is  given by,
\begin{eqnarray}
T_{\mu\nu}^\psi = - \eta_{\mu\nu}({\overline \psi} i \gamma^{\sigma}D_{\sigma} \psi - m_\psi {\overline \psi} \psi)   
+ \frac{i}{2}\left[ ( {\overline \psi} \gamma_{\mu}D_{\nu} \psi  + {\overline \psi} \gamma_{\nu}D_{\mu} \psi )\right] 
+ \frac{i}{2} \eta_{\mu\nu} \partial_{\sigma}({\overline \psi}  \gamma^{\sigma}\psi) 
\nonumber \\ 
- \frac{i}{4}\left[\partial_{\mu}({\overline \psi}  \gamma_{\nu} \psi) + \partial_{\nu}({\overline \psi}  \gamma_{\mu} \psi) \right]
\label{lf1}
\end{eqnarray} 
Now taking the functional derivative of ${\mathcal{L}}_\psi(\hat g)$ successively, one finds
that
\begin{eqnarray}
\nonumber
\left. \frac { \delta^{2}
\mathcal{L}_{\psi} }
{ \delta\hat g^{\mu\nu}  \delta\hat g^{\rho \sigma}}
\right|_{\hat g = \eta} & = & - \frac{1}{128}[\eta_{\mu \rho}
\partial^{\alpha}({\overline \psi}\epsilon_{\nu\sigma\alpha\tau}\gamma^{\tau}\gamma^5 \psi)
\nonumber \\
&+& \eta_{\mu \sigma}
\partial^{\alpha}({\overline \psi}\epsilon_{\nu\rho\alpha\tau}\gamma^{\tau}\gamma^5 \psi) +
\eta_{\nu \sigma}
\partial^{\alpha}({\overline \psi}\epsilon_{\mu\rho\alpha\tau}\gamma^{\tau}\gamma^5 \psi) +
\eta_{\nu \rho}
\partial^{\alpha}({\overline \psi}\epsilon_{\mu\sigma\alpha\tau}\gamma^{\tau}\gamma^5 \psi)]                                                
\label{lf2}
\end{eqnarray}
After substituting Eq. (\ref{lf}),~(\ref{lf1}) and (\ref{lf2}) in the 
expression of ${\mathcal{L}}^{mn}_2$ (Eq. (\ref{la})), we get the Feynman
rule
for the vertex involving two fermions and two gravitons as follows,

\begin{eqnarray}  
 h_{\mu\nu}^{(n)}  h_{\rho\sigma}^{(m)}\psi\psi: \frac{i}
{4\Lambda_\pi^2}
\delta_{MN}[[ \eta_{\sigma\mu}(\gamma_\rho (p_1 + p_2)_\nu 
+ \gamma_\nu (p_1 + p_2)_\rho - 2 \eta_{\rho\nu}(\slash{p_1} +
\slash{p_2} 
- 2 m_\psi ) )] 
\nonumber \\
 +  [\eta_{\rho\mu}(\gamma_\sigma
(p_1 + p_2)_\nu 
+ \gamma_\nu (p_1 + p_2)_\sigma - 2 \eta_{\sigma\nu}(\slash{p_1} +
\slash{p_2} 
- 2 m_\psi ) )]
\nonumber \\
+  [\eta_{\nu\rho}(\gamma_\mu (p_1 +
p_2)_\sigma 
+ \gamma_\sigma (p_1 + p_2)_\mu - 2 \eta_{\mu\sigma}(\slash{p_1} +
\slash{p_2} - 2 m_\psi ) )]
\nonumber \\
+  [\eta_{\nu\sigma}(\gamma_\mu (p_1
+ p_2)_\rho + \gamma_\rho (p_1 + p_2)_\mu - 2 \eta_{\mu\rho}(\slash{p_1}
+ \slash{p_2} - 2 m_\psi ) )]
\nonumber \\
-  [\eta_{\mu\nu} (\gamma_\rho (p_1+
p_2)_\sigma + \gamma_\sigma (p_1 + p_2)_\rho - 2 \eta_{\rho\sigma}
(\slash{p_1} + \slash{p_2} - 2 m_\psi ) )]
\nonumber \\
-  [\eta_{\rho\sigma} (\gamma_\mu
(p_1 + p_2)_\nu + \gamma_\nu (p_1 + p_2)_\mu - 2 \eta_{\mu\nu}
(\slash{p_1} + \slash{p_2} - 2 m_\psi ) )]
\nonumber \\
+ 2 C_{\mu\nu,~\rho\sigma} [
({\slash{p}}_1 + \slash{p_2} - 2 m_\psi )]
+ \frac{1}{2}[\eta_{\mu \rho}
\epsilon_{\nu\sigma\lambda\tau}\gamma^{\tau}\gamma^5 (p_1 -
p_2)^{\lambda} 
\nonumber \\
+ \eta_{\mu \sigma}
\epsilon_{\nu\rho\lambda\tau}\gamma^{\tau}\gamma^5 (p_1 -
p_2)^{\lambda} + \eta_{\nu \sigma}
\epsilon_{\mu\rho\lambda\tau}\gamma^{\tau}\gamma^5 (p_1 -
p_2)^{\lambda} + \eta_{\nu \rho}
\epsilon_{\mu\sigma\lambda\tau}\gamma^{\tau}\gamma^5 (p_1 -
p_2)^{\lambda} ]]
\end{eqnarray}  
where $\gamma^5 = \gamma^0 \gamma^1 \gamma^2 \gamma^3$ and $\delta_{MN}$
is defined in the flavour basis (M, N are the flavour indices) and $p_1$,~$p_2$
are the four momenta of the two fermions.

Similarly the Feynman rules for the vertices involving two gravitons,
two fermions and one gauge particle is,

\begin{eqnarray}
h_{\mu\nu}^{(n)}  h_{\rho\sigma}^{(m)}A^a_{\lambda} \psi\psi
: \frac{i g}{\Lambda_\pi^2} T^{a}_{MN}\left[\gamma^{\eta}
\left[\eta_{\nu\sigma}
(C_{\mu\rho,~\lambda\eta} - \eta_{\mu\rho}\eta_{\lambda\eta}) + (\nu
\leftrightarrow \mu)\right]\noindent
\right.
\nonumber \\
+ \gamma^{\eta}\left[\{ \eta_{\nu\rho}
(C_{\mu\sigma,~\lambda\eta} 
- \eta_{\mu\sigma}\eta_{\lambda\eta})
+(\nu \leftrightarrow \mu)\}\right]\noindent
\nonumber \\
+ \gamma^{\eta}\left[\eta_{\sigma\rho}
(C_{\mu\nu,~\lambda\eta} - \eta_{\mu\nu}\eta_{\lambda\eta}) 
+ \eta_{\mu\nu} (C_{\rho\sigma,~\lambda\eta} -
\eta_{\rho\sigma}\eta_{\lambda\eta}) \right] 
\nonumber \\
\left.
- \gamma^{\eta}\left[\eta_{\lambda\eta}
(C_{\mu\nu,~\rho\sigma}) \right]\right]
\end{eqnarray}

\vskip 10pt
\noindent
{\large\bf {B.3 Coupling To Gauge Fields}}

\vspace{.5cm}

The lagrangian and conserved energy-momentum tensor for the gauge field
are given by,
\begin{eqnarray}
{\mathcal{L}}_A(\hat {g}) = - \frac{1}{4}F^{\tau\eta}F_{\tau\eta} +
\frac{1}{2} {m_A}^2 A^
{\tau}A_{\tau} - \frac{1}{2 \xi} (\partial_\tau A^\tau +
\Gamma^{\tau}_{\tau \lambda} A^\lambda)^2
\label{lg}
\end{eqnarray}

\begin{eqnarray}
T_{\mu\nu} = \eta_{\mu\nu}\left[ \frac{1}{4}F^{\tau\eta}F_{\tau\eta} - \frac{1}{2} {m_A}^2 A^{\tau}A_{\tau}\right] - \left[F_\mu^\tau F_{\nu\tau} - m_A^2 A_\mu A_\nu\right]  + \frac{1}{\xi}\left[\partial_\mu \partial^\tau A_\tau A_\nu + \partial_\nu \partial^\tau A_\tau A_\mu\right] 
\nonumber \\
-\frac{1}{\xi} \eta_{\mu\nu}\left[ \partial_\eta \partial^\tau A_\tau A^\eta + \frac{1}{2} (\partial^\tau A_\tau)^2\right] 
\label{lg1}
\end{eqnarray}

After successive functional differentiation, one obtain,
\begin{eqnarray}
\left. \frac { \delta^{2}
\mathcal{L}_{\psi} }
{ \delta\hat g^{\mu\nu}  \delta\hat g^{\rho \sigma}}
\right|_{\hat g = \eta} & = & 
- \frac{1}{4}\left[F_{\mu\sigma}F_{\nu\rho} +  
F_{\mu\rho}F_{\nu\sigma}\right]_{\hat g = \eta} + \frac{1}{4 \xi}\left[\partial_{\rho}\partial_{\mu}A_{\nu}A_{\sigma} + 
\partial_{\rho}\partial_{\nu}A_{\mu}A_{\sigma}\right]
\nonumber \\
& &
+\frac{1}{4 \xi}\left[\partial_{\sigma}\partial_{\mu}A_{\nu}A_{\rho} 
+\partial_{\sigma}\partial_{\nu}A_{\mu}A_{\rho}\right] 
- \frac{1}{4 \xi}\eta_{\rho\sigma}\left[(\partial_\mu A_\nu + \partial_\nu A_\mu) \partial_\eta A^{\eta}\right]  
\nonumber \\
& &
- \frac{1}{4 \xi}\eta_{\rho\sigma}\left[(\partial_\eta \partial_\mu A_\nu A^{\eta}
+ \partial_\eta \partial_\nu A_\mu A^{\eta}) \right]  
\nonumber \\
& &
+ \frac{1}{4 \xi} C_{\mu\nu,\rho\sigma}\left[(\partial_\tau \partial_\eta A^\eta A^{\tau} 
+ (\partial_{\tau}A^{\tau})^2 \right]
\label{lg2}
\end{eqnarray}

The sets of Eq. (\ref{lg}),~(\ref{lg1}) and (\ref{lg2}) together when
substituted in 
Eq. (\ref{la}), gives us the 
Feynman rule for the vertex involving two gravitons and two gauge fields
as follows:
\begin{eqnarray}
 h_{\mu\nu}^{(n)} 
h_{\rho\sigma}^{(m)}A^a_{\tau}A^b_{\eta}:
\frac{i}{ \Lambda_\pi^2} \delta^{ab}\left[\left[ \eta_{\sigma\mu}(
C_{\rho\nu,~\tau\eta} (k_1.k_2 + m_A^2) + D_{\rho\nu,~\tau\eta}(k_1,k_2) +
\xi^{-1} E_{\rho\nu,~\tau\eta}) + (\rho \leftrightarrow \sigma)\right]
\right.
\nonumber \\
+\left[ \eta_{\nu\rho}(
C_{\sigma\mu,~\tau\eta} (k_1.k_2 + m_A^2) +
D_{\sigma\mu,~\tau\eta}(k_1,k_2) + \xi^{-1} E_{\sigma\mu,~\tau\eta}) +
(\rho \leftrightarrow \sigma)\right] \nonumber \\
- \left[ \eta_{\mu\nu}(
C_{\rho\sigma,~\tau\eta}(k_1.k_2 + m_A^2) +
D_{\rho\sigma,~\tau\eta}(k_1,k_2) + \xi^{-1} E_{\rho\sigma,~\tau\eta} ) +
(\mu \leftrightarrow \rho , \nu \leftrightarrow \sigma)\right]
\nonumber \\
+  C_{\mu\nu,~\rho\sigma}\left[ (k_1.k_2 +
m_A^2)\eta_{\tau\eta} - (k_{1\eta}k_{2\tau} - \xi^{-1}k_{1\tau}k_{2\eta})\right]
\nonumber \\
+ \left[ (\eta_{\rho\tau}\eta_{\sigma\eta}k_{1\nu}k_{2\mu} + \eta_{\rho\eta}\eta_{\tau\sigma}k_{1\mu}k_{2\nu})  
- (\eta_{\rho\tau}\eta_{\mu\eta}k_{1\nu}k_{2\sigma} + \eta_{\rho\eta}\eta_{\tau\mu}k_{1\sigma}k_{2\nu}) + (\rho \leftrightarrow \sigma)\right]  
\nonumber \\
+ \left[ (\eta_{\nu\tau}\eta_{\mu\eta}k_{1\rho}k_{2\sigma} + \eta_{\nu\eta}\eta_{\tau\mu}k_{1\sigma}k_{2\rho})  
- (\eta_{\nu\tau}\eta_{\sigma\eta}k_{1\rho}k_{2\mu} + \eta_{\nu\eta}\eta_{\tau\sigma}k_{1\mu}k_{2\rho}) + (\rho \leftrightarrow \sigma)\right]  
\nonumber \\
- \frac{1}{\xi}\left[ \eta_{\nu\tau}\eta_{\sigma\eta}(k_{1\mu}k_{1\rho} +
k_{2\mu}k_{2\rho})+(\mu \leftrightarrow \nu)\right]
- \frac{1}{\xi}\left[ \eta_{\nu\tau}\eta_{\rho\eta}(k_{1\mu}k_{1\sigma} +
k_{2\mu}k_{2\sigma})+ (\mu \leftrightarrow \nu)\right]
\nonumber \\
+ \frac{1}{\xi}\left[ \eta_{\nu\tau}\eta_{\rho\sigma}(k_{1\eta}k_{2\mu} +
k_{1\mu}k_{2\eta})+ (\mu \leftrightarrow \nu)\right]
+ \frac{1}{\xi}\left[ \eta_{\nu\tau}\eta_{\rho\sigma}(k_{1\eta}k_{1\mu} +
k_{2\mu}k_{2\eta})+ (\mu \leftrightarrow \nu)\right]
\nonumber \\
\left.
- \frac{1}{\xi}\left[
C_{\mu\nu,~\rho\sigma}(k_{1\eta}k_{1\tau} + k_{2\tau}k_{2\eta})+ (\mu \leftrightarrow \nu)\right]
- \frac{1}{\xi}\left[2 C_{\mu\nu,~\rho\sigma}(
k_{1\tau}k_{2\eta}) \right]\right]
\end{eqnarray}
where $k_1$ and $k_2$ are the four-momenta of the gauge
fields $A_{\tau}$ and $A_{\eta}$.

Proceeding in the same way, one also finds the Feynman rules for the
vertices involving two gravitons, three gauge bosons and two gravitons, four
gauge bosons $$\left[ h_{\mu\nu}^{(n)}
h_{\rho\sigma}^{(m)}A^a_{\lambda}(k_1)A^b_{\eta}(k_2)A^c_{\delta}(k_3)\right]$$
as follows,
\begin{eqnarray}
- \frac{ig}{\Lambda_\pi^2} f^{abc} [ \large [\{ \eta_{\sigma \nu}
C_{\mu\rho,~\lambda\eta}
(k_1 - k_2)_\delta + \eta_{\sigma \nu} C_{\mu\rho,~\lambda\delta}(k_3 -
k_1)_\eta 
+  \eta_{\sigma \nu}  C_{\mu\rho,~\eta\delta} (k_2 - k_3)_\lambda 
+ (\mu \leftrightarrow \nu ) \}
\nonumber \\ 
+\{ (\rho \leftrightarrow \sigma) \} \large ]
+\{ \eta_{\sigma\nu}F_{\mu\rho,~\lambda\eta\delta} 
(k_1,k_2,k_3) + \eta_{\sigma\mu}F_{\nu\rho,~\lambda\eta\delta}(k_1,k_2,k_3) 
+ (\rho \leftrightarrow \sigma)\}  
\nonumber \\  
- \eta_{\rho\sigma} C_{\mu\nu,~\lambda\eta}(k_1 - k_2)_\delta 
-  \eta_{\rho\sigma} C_{\mu\nu,~\lambda\delta}(k_3 - k_1)_\eta 
- \eta_{\rho\sigma} C_{\mu\nu,~\eta\delta}(k_2 - k_3)_\lambda
\nonumber \\
-  \eta_{\mu\nu} C_{\rho\sigma,~\lambda\eta}(k_1 - k_2)_\delta
-  \eta_{\mu\nu} C_{\rho\sigma,~\lambda\delta}(k_3 - k_1)_\eta
-  \eta_{\mu\nu} C_{\rho\sigma,~\eta\delta}(k_2 - k_3)_\lambda
\nonumber \\
- \eta_{\rho\sigma} F_{\mu\nu,~\lambda\eta\delta}(k_1,k_2,k_3)
- \eta_{\mu\nu} F_{\rho\sigma,~\lambda\eta\delta}(k_1,k_2,k_3)
\nonumber \\
+ C_{\mu\nu,~\rho\sigma} ( \eta_{\lambda\eta} (k_1 - k_2)_\delta 
+ \eta_{\eta\delta} (k_2 - k_3)_\lambda + \eta_{\lambda\delta}
(k_3 - k_1)_\eta )
\nonumber \\
- \large [ \{ (k_{3 \mu} \eta_{\delta\sigma} - 
k_{3\sigma} \eta_{\mu\delta})(\eta_{\rho\eta}\eta_{\nu\lambda} 
- \eta_{\nu\eta}\eta_{\rho\lambda} ) 
+ \eta_{\nu\delta}(\eta_{\sigma\eta}\eta_{\rho\lambda}k_{2\mu}
- \eta_{\rho\eta}\eta_{\sigma\lambda}k_{1\mu}
+ \eta_{\rho\eta}\eta_{\mu\lambda}k_{1\sigma} 
- \eta_{\rho\lambda}\eta_{\mu\eta}k_{2\sigma} )
\nonumber \\
+  \eta_{\rho\delta}(\eta_{\sigma\lambda}\eta_{\nu\eta}k_{2\mu}
- \eta_{\sigma\eta}\eta_{\nu\lambda}k_{1\mu}
+ \eta_{\mu\eta}\eta_{\nu\lambda}k_{1\sigma}
- \eta_{\mu\lambda}\eta_{\nu\eta}k_{2\sigma})
+ (\rho \leftrightarrow \sigma ) \}
- (\nu \leftrightarrow \mu , \rho \leftrightarrow \sigma)
]\large ]
\end{eqnarray}
Finally the Feynman rule for the vertices involving two gravitons and 
four gauge bosons$$\left[
h_{\mu\nu}^{(n)} h_{\rho\sigma}^{(m)}
A^a_{\lambda}(k_1)A^b_{\eta}(k_2)A^c_{\delta}(k_3)A^d_{\tau}(k_4)\right]$$
are given by,
\bea
\frac{i g^2}{\Lambda_\pi^2} [ \{\eta_{\sigma\nu}(f^{eac}f^{ebd}
G_{\mu\rho,~\lambda\eta\delta\tau} + f^{eab}f^{ecd}
G_{\mu\rho,~\lambda\delta\eta\tau} + f^{ead}f^{ebc}
G_{\mu\rho,~\lambda\eta\tau\delta}) + (\mu \leftrightarrow \nu )\}
\nonumber \\ 
+ \{\eta_{\nu\rho}(f^{eac}f^{ebd}
G_{\mu\sigma,~\lambda\eta\delta\tau} + f^{eab}f^{ecd}
G_{\mu\sigma,~\lambda\delta\eta\tau} + f^{ead}f^{ebc}
G_{\mu\sigma,~\lambda\eta\tau\delta}) + (\mu \leftrightarrow \nu )\} 
\nonumber \\
- \eta_{\rho\sigma}(f^{eac}f^{ebd}G_{\mu\nu,~\lambda\eta\delta\tau}
+  f^{eab}f^{ecd}G_{\mu\nu,~\lambda\delta\eta\tau} + f^{ead}f^{ebc}
G_{\mu\nu,~\lambda\eta\tau\delta})\noindent
\nonumber \\
- \eta_{\mu\nu}(f^{eac}f^{ebd}G_{\rho\sigma,~\lambda\eta\delta\tau}
+  f^{eab}f^{ecd}G_{\rho\sigma,~\lambda\delta\eta\tau} + f^{ead}f^{ebc}
G_{\rho\sigma,~\lambda\eta\tau\delta})\noindent
\nonumber \\
- C_{\mu\nu,~\rho\sigma}\{
f^{eac}f^{ebd}(\eta_{\lambda\eta}\eta_{\delta\tau}
- \eta_{\lambda\tau} \eta_{\delta\eta} ) +  f^{eab}f^{ecd}(\eta_{\lambda\delta} 
\eta_{\eta\tau} - \eta_{\lambda\tau}\eta_{\delta\eta} ) 
+  f^{ead}f^{ebc} (\eta_{\lambda\eta} \eta_{\tau\delta} - \eta_{\lambda\delta}
\eta_{\tau\eta}) \}
\nonumber \\
- [ \{f^{eab}f^{ecd}( \eta_{\nu\eta}\eta_{\sigma\lambda}\eta_{\rho\delta} -
\eta_{\rho\delta}\eta_{\nu\lambda}\eta_{\sigma\eta}) 
+ f^{eca}f^{ebd}(\eta_{\rho\eta} \eta_{\nu\lambda} \eta_{\delta\sigma} - \eta_{\rho\eta}
\eta_{\nu\delta} \eta_{\sigma\lambda}) 
\nonumber \\
+ f^{ebc} f^{eda}(\eta_{\rho\lambda}
\eta_{\nu\eta} \eta_{\delta\sigma} - \eta_{\rho\lambda} \eta_{\sigma\eta}
\eta_{\nu\delta})\} \eta_{\mu\tau}
+ \{ f^{eab}f^{ecd} (\eta_{\mu\delta} \eta_{\nu\lambda} \eta_{\sigma\eta}
- \eta_{\mu\delta}\eta_{\nu\eta}  \eta_{\sigma\lambda})
\nonumber \\
+  f^{eca}f^{ebd} (\eta_{\nu\delta}\eta_{\mu\eta} \eta_{\sigma\lambda}
- \eta_{\mu\eta} \eta_{\nu\lambda}  \eta_{\delta\sigma})
+ f^{ebc} f^{eda} (\eta_{\mu\lambda}\eta_{\nu\eta} \eta_{\delta\sigma}
- \eta_{\mu\lambda} \eta_{\sigma\eta} \eta_{\nu\delta})  \} \eta_{\rho\tau}
\nonumber \\
+ \{ f^{eab}f^{ecd} (\eta_{\delta\sigma}\eta_{\mu\eta} \eta_{\rho\lambda}-
 \eta_{\delta\sigma}\eta_{\mu\lambda}\eta_{\rho\eta}) +  f^{eca}f^{ebd}
( \eta_{\rho\delta} \eta_{\mu\lambda}\eta_{\sigma\eta} - \eta_{\mu\delta}
 \eta_{\rho\lambda}\eta_{\sigma\eta})
\nonumber \\
+ f^{ebc} f^{eda} (\eta_{\rho\delta} \eta_{\mu\eta} \eta_{\sigma\lambda}
- \eta_{\mu\delta} \eta_{\rho\eta}\eta_{\sigma\lambda}) \} \eta_{\nu\tau}
\nonumber \\
+ \{ f^{eab}f^{ecd} (\eta_{\nu\delta}\eta_{\mu\lambda}\eta_{\rho\eta} -
\eta_{\nu\delta}\eta_{\mu\eta}\eta_{\rho\lambda})
+  f^{eca}f^{ebd} (\eta_{\mu\delta}\eta_{\rho\lambda}\eta_{\nu\eta}-
\eta_{\rho\delta}\eta_{\mu\lambda}\eta_{\nu\eta})
\nonumber \\
+ f^{ebc} f^{eda}( \eta_{\mu\delta}\eta_{\rho\eta}\eta_{\nu\lambda}
- \eta_{\rho\delta}\eta_{\mu\eta}\eta_{\nu\lambda}) \} \eta_{\tau\delta}
+ \{\rho \leftrightarrow \sigma\} ] 
\eea 
where, 
$$
D_{\mu\nu,~ \rho\sigma(k_1, k_2)} = \eta_{\mu\nu}k_{1\sigma}k_{2 \rho}
- [  \eta_{\mu\sigma}k_{1\nu}k_{2 \rho} +  \eta_{\mu\rho} k_{1\sigma}
k_{2\nu} -  \eta_{ \rho\sigma}k_{1\mu}k_{2\nu} 
+ (\mu \leftrightarrow \nu)]   
$$
\vspace{-0.15in}
$$
E_{\mu\nu,~ \rho\sigma(k_1, k_2)} = \eta_{\mu\nu}(k_{1\rho}k_{1
\sigma}
+ k_{2\rho}k_{2 \sigma} + k_{1\rho}k_{2 \sigma})
- [\eta_{\nu\sigma}k_{1\mu}k_{1\rho} + \eta_{\nu\rho} k_{2\mu}k_{2 \sigma}
+ (\mu \leftrightarrow \nu)]
$$
\vspace{-0.15in}
$$
F_{\mu\nu,~ \rho\sigma\lambda(k_1,k_2,k_3)} = \eta_{\mu\rho}  \eta_{
\sigma\lambda}
(k_2 - k_3)_{\nu} +  \eta_{\mu\sigma}  \eta_{\rho\lambda} (k_3 - k_1)_{\nu}
+ \eta_{\mu\lambda}  \eta_{\rho\sigma} (k_1 - k_2)_{\nu} + (\mu \leftrightarrow \nu)
$$
\vspace{-0.15in}
$$
G_{\mu\nu,~ \rho\sigma\lambda\delta} =  \eta_{\mu\nu}(\eta_{\rho\sigma} 
\eta_{\lambda \delta} - \eta_{\rho\delta} \eta_{\sigma \lambda}) 
+ \left[ \eta_{\mu\rho}  \eta_{\nu \delta}  \eta_{\lambda \sigma}
+ \eta_{\mu\lambda}  \eta_{\nu \sigma} \eta_{\rho  \delta} -  \eta_{\mu \rho}
 \eta_{\nu \sigma}  \eta_{\lambda  \delta} -  \eta_{\mu\lambda} \eta_{\nu \delta} 
\eta_{\rho\sigma}\right] 
\nonumber \\
+\left[ \mu \leftrightarrow \nu \right] $$ 

\bigskip
\noindent
{\Large\bf Appendix B: Triple graviton vertex}\\

To find the feynman rules for three graviton vertex we 
expand the pure gravity Lagrangian upto next order 
in weak field approximation. In this approximation 
\begin{eqnarray}
\nonumber
L_{g} &=& \frac{1}{\kappa^2} \sqrt{g} R = L_{0} + \kappa L_{1} + ...\\
\nonumber 
&=& \frac{1}{4} \left(\partial^{\mu} h^{\nu\rho}\partial_{\mu} h_{\nu\rho} 
- \partial^{\mu} h_{\nu}^{\nu}\partial_{\mu} h_{\rho}^{\rho} - \partial^{\mu}
h_{\mu}^{\nu}\partial_{\rho} h^{\rho\nu} + 2 \partial_{\mu}
h^{\mu\nu}\partial_{\nu} h_{\rho}^{\rho}\right)
\\
\nonumber
&-& \frac{\kappa}{4}\left[h^{\mu\nu}\partial_{\mu}
h^{\alpha\beta}\partial_{\nu} h_{\alpha\beta} -
\frac{1}{2}h^{\mu\nu}\partial_{\mu}
h^{\alpha\alpha}\partial_{\nu} h_{\beta\beta}\right]  
\\
&-&\frac{\kappa}{4}\left[2 h^{\mu\nu}\partial^{\beta}
h_{\mu\alpha}\partial^{\alpha} h_{\nu\beta} + h^{\mu\nu}\partial_{\alpha}
h^{\beta\beta}\partial^{\alpha} h_{\mu\nu} - 2
h^{\mu\nu}\partial_{\alpha}
h_{\mu\beta}\partial^{\alpha} h^{\beta}_{\nu}\right] + ....
\end{eqnarray}
From the expression given in $L_1$, feynman rule for three graviton
vertex reads
\begin{eqnarray}
\nonumber
&&\frac{\kappa}{8}\left[ (k_{2\mu}k_{3\nu} +
k_{2\nu}k_{3\mu})F_{\rho\lambda k\sigma }  + 
 (k_{1\rho} k_{3\sigma} +
k_{1\sigma} k_{3\rho}) F_{\mu\lambda k \nu}  + (k_{1\lambda}k_{2 k} +
k_{1 k}k_{2\lambda}) F_{\mu\rho\sigma\nu} \right] -
\\
\nonumber
&&
\frac{\kappa}{8}\left[ (k_{2\mu}k_{3\nu} +
k_{2\nu}k_{3\mu})\eta_{\lambda k}\eta_{\rho \sigma} + \eta_{\mu
\nu} \eta_{\lambda k}(k_{1\rho}k_{3\sigma} + k_{1\sigma}k_{3\rho})\right]-
\\
\nonumber
&&
\frac{\kappa}{8}\left[  \eta_{\mu
\nu} \eta_{\rho \sigma}(k_{1\lambda}k_{2 k} +
k_{1 k}k_{2\lambda})  \right] + 
\\
\nonumber
&&
\frac{\kappa}{8}\left[F_{\mu\rho\lambda\nu} k_{2k} k_{3\sigma}
+ F_{\mu\rho k\nu} k_{2\lambda} k_{3\sigma} + F_{\mu\sigma\lambda\nu}
k_{2k}
k_{3\rho}+F_{\mu\sigma k \nu} k_{2 \lambda} k_{3\rho}       
\right] +
\\
\nonumber
&&
\frac{\kappa}{8}\left[F_{\mu\rho\sigma\lambda} k_{1k} k_{3\nu}+
F_{\mu\rho\sigma k} k_{1\lambda} k_{3\nu} + F_{\nu\rho\sigma\lambda}
k_{1k} k_{3\mu} + F_{\nu\rho\sigma k} k_{1\lambda}
k_{3\mu}\right] + 
\\
\nonumber
&&
\frac{\kappa}{8}\left[F_{\mu\lambda k \rho} k_{1\sigma} k_{2\nu} +
F_{\mu\lambda k \sigma}
k_{1\rho} k_{2\nu} + F_{\nu\lambda k \rho} k_{1\sigma}
k_{2\nu}+F_{\nu\lambda k \sigma} k_{1\rho} k_{2\mu}\right] +
\\
\nonumber
&&
\frac{\kappa}{8}\left[G_{\mu\nu\rho\lambda\sigma k} k_{1}.k_{2}
+ G_{\rho\sigma\mu\lambda\nu k} k_{2}.k_{3} +
G_{\mu\nu\lambda\rho k \sigma} k_{1}.k_{3}\right] -
\\
&& \frac{\kappa}{8}\left[I_{\mu\nu\rho\sigma\lambda k}(k_1.k_2 
+ k_2.k_3 + k_3.k_1)\right]            
\end{eqnarray}
where F, G and I are defined by the following relations 
\begin{eqnarray}
F_{\mu\nu\rho\sigma} = \eta_{\mu\nu}\eta_{\rho\sigma} +
\eta_{\mu\rho}\eta_{\nu\sigma} \\
G_{\mu\nu\rho\lambda\sigma k} = \eta_{\mu\nu} F_{\rho\lambda k \sigma}
+ \eta_{\rho\sigma} F_{\mu\lambda k \nu}\\
I_{\mu\nu\rho\sigma\lambda k} = \eta_{\mu\rho} F_{\nu\lambda k \sigma}
+ \eta_{\mu\sigma} F_{\nu\lambda k \rho} + \eta_{\nu\rho} F_{\mu\lambda k
\sigma} + \eta_{\nu\sigma}  F_{\mu\lambda k \rho}  
\end{eqnarray}
\begin{figure}[tb]
\hbox{\hspace{0em}
\hbox{\psfig{figure=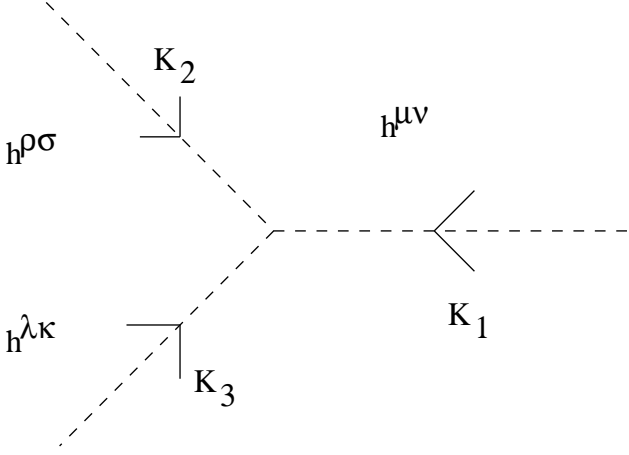,height=6cm}}}
\caption {A Typical Triple Graviton Vertex.}
\label{fig-3}
\end{figure}

\bigskip
\noindent
{\large \bf Appendix C: Cross section for two graviton production through
fermion anti-fermion annihilation}\\

In this Appendix we give the detailed form of the cross section for 
two graviton production by fermion anti-fermion annihilation. The
differential cross section for this process is given by. 
\begin{eqnarray}
\frac{d\sigma}{d(cos\theta)}=\frac{1}{32\pi
s}\sqrt{1 - \frac{4 M_n^2}{s}}(T_{11}+T_{22}+T_{33}+T_{12}
+T_{23}+T_{13}+T_{44}+T_{14}+T_{24}+T_{34})
\label{Eq:ffbartoGG}
\end{eqnarray}
where the terms $T_{ij}$  inside bracket are given by,
\begin{equation}
\nonumber
T_{11} = \frac{\pi^2 c_0^4}{m_0^4}\left(\frac{-2 {\left( {M_n}^2 - t
\right) }^4 
    \left( 9 {M_n}^8 - 4 t^3 u + 
      12{M_n}^2 t^2 \left( t + u \right)  - 
      {M_n}^4 t \left( 20 t + 9 u \right) 
      \right) }{9 {M_n}^8 t^2} \right) 
\end{equation}
\begin{equation}
T_{22} = \frac{\pi^2 c_0^4}{m_0^4}\left(\frac{-2{\left( {M_n}^2 - u
\right) }^4
    \left( 9 {M_n}^8 - 4 t u^3 + 
      12 {M_n}^2 u^2 \left( t + u \right)  - 
      {M_n}^4 u \left( 9 t + 20 u \right) 
      \right) }{9 {M_n}^8 u^2} \right)
\end{equation}
\begin{eqnarray}
\nonumber
T_{33} &=& \frac{\pi^2 c_0^4}{m_0^4}\left(\frac{2 \left(
1092{M_n}^{12} - 
      1230 {M_n}^{10}
       \left( t + u \right)\right)  + 
      16 t {\left( t - u \right) }^2 u 
       {\left( t + u \right) }^2}{9 {M_n}^8}\right) -
\\
\nonumber 
&&      
\frac{\pi^2 c_0^4}{m_0^4} \left(\frac{7 {M_n}^6\left( t + u \right) 
       \left( 25 t^2 + 52 t u + 25 u^2 \right)}{9 {M_n}^8}\right)  + 
\\
\nonumber
&&  
    \frac{\pi^2 c_0^4}{m_0^4} \left(\frac{4 {M_n}^8 
       \left( 113 t^2 + 299 t u + 113 u^2 \right)}{9 {M_n}^8}\right)  - 
\\
\nonumber 
&&       
\frac{\pi^2 c_0^4}{m_0^4}\left(\frac{{M_n}^2 \left( t + u \right) 
       \left( 15 t^4 + 62 t^3 u - 34 t^2 u^2 + 
         62 t u^3 + 15 u^4 \right)}{9 {M_n}^8}\right)  + 
\\
&&     
  \frac{\pi^2 c_0^4}{m_0^4}\left(\frac{{M_n}^4
       \left( 56 t^4 + 370 t^3 u + 84 t^2 u^2 + 
         370 t u^3 + 56 u^4 \right)}{9{M_n}^8}\right) 
\end{eqnarray}
\begin{eqnarray}
\nonumber
T_{12} &=&  \frac{\pi^2 c_0^4}{m_0^4}\left( \frac {\left( 2 {M_n}^2 - t - u \right) 
    \left( 88 {M_n}^{14} - 62 {M_n}^{12} \left( t + u \right)  - 
      4 t^3 u^3 \left( t + u \right)\right)}{9 M_n^8 tu}\right)  - 
\\
\nonumber      
&&  \frac{\pi^2 c_0^4}{m_0^4}\left(\frac{\left( 2 {M_n}^2 - t - u \right)\left(10 {M_n}^{10}
       \left( 7 t^2 + 19 t u + 7 u^2 \right)\right)}{9 M_n^8 tu}\right)  - 
\\
\nonumber
&&      
   \frac{\pi^2 c_0^4}{m_0^4}\left(\frac{\left( 2 {M_n}^2 - t - u \right)2 {M_n}^2 t^2 u^2 
       \left( 9 t^2 + 14 t u + 9 u^2 \right)}{9 M_n^8 tu}\right)  -
\\
\nonumber
&& 
       \frac{\pi^2 c_0^4}{m_0^4}\left(\frac{\left( 2 {M_n}^2 - t - u \right)\left({M_n}^4 t u
\left( t
+ u \right) 
       \left( 26 t^2 + 47 t u + 26 u^2 \right)\right)}{9 M_n^8 tu}\right)  + 
\\
\nonumber
&&      
 \frac{\pi^2 c_0^4}{m_0^4}\left(\frac{\left( 2 {M_n}^2 - t - u \right)\left({M_n}^8 \left( t + u
\right)
       \left( 62 t^2 + 41 t u + 62 u^2 \right)\right)}{9 M_n^8 tu}\right)  + 
\\
&&     
  \frac{\pi^2 c_0^4}{m_0^4}\left(\frac{\left( 2 {M_n}^2 - t - u \right)\left({M_n}^6 \left( -12 t^4 + 64 t^3 u + 
         202 t^2 u^2 + 64 t u^3 - 12 u^4 \right)  \right) }
    {9 {M_n}^8 t u}\right)
\end{eqnarray}
\begin{eqnarray}
\nonumber
T_{13} &=& \frac{\pi^2 c_0^4}{m_0^4}\left(\frac{8 \left( -99 {M_n}^{14} + 
      2 t^4 \left( t - u \right) u
       \left( t + u \right)  + 
      {M_n}^{12} \left( 96 t + 63 u \right)\right)}{9 {M_n}^8 t}\right)  + 
\\
\nonumber      
&&\frac{\pi^2 c_0^4}{m_0^4}\left(\frac{8\left({M_n}^{10}
       \left( 5 t^2 + 70 t u - 9 u^2 \right)  - 
      {M_n}^8 t 
       \left( 3 t^2 + 92 t u + 55 u^2 \right)\right)}{9 {M_n}^8
t}\right)  - 
\\      
\nonumber
&&\frac{\pi^2 c_0^4}{m_0^4}\left(\frac{8\left({M_n}^2 t^3 
       \left( 3 t^3 + 12 t^2 u - 5 t u^2 + 2 u^3 \right)
          + {M_n}^6 t 
       \left( -18 t^3 - 40 t^2 u + t u^2 + 6 u^3 \right)\right)}{9 {M_n}^8 
t}\right) +
\\ 
&&\frac{\pi^2 c_0^4}{m_0^4}\left(\frac{8\left({M_n}^4 t^2 
       \left( 13 t^3 + 36 t^2 u + 31 t u^2 + 7 u^3
         \right)  \right)}{9 {M_n}^8 t}\right) 
\end{eqnarray}
\begin{eqnarray}
\nonumber
T_{23} &=& \frac{\pi^2 c_0^4}{m_0^4}\left(\frac{8 \left( -99 {M_n}^{14} - 
      2 t \left( t - u \right) u^4
       \left( t + u \right)  + 
      {M_n}^{12} \left( 63 t + 96 u \right)\right)}{9 {M_n}^8 u}\right) - 
\\
\nonumber      
&&\frac{\pi^2 c_0^4}{m_0^4}\left(\frac{8\left({M_n}^8 u
       \left( 55 t^2 + 92 t u + 3 u^2 \right)  + 
      {M_n}^{10}
       \left( -9 t^2 + 70 t u + 5 u^2 \right)\right)}{9 {M_n}^8
u}\right)  + 
\\
\nonumber      
&&\frac{\pi^2 c_0^4}{m_0^4}\left(\frac{8\left({M_n}^6 u
       \left( 6 t^3 + t^2 u - 40 t u^2 - 18 u^3 \right) 
       - {M_n}^2 u^3 
       \left( 2 t^3 - 5 t^2 u + 12 t u^2 + 3 u^3 \right)\right)}{9 
{M_n}^8 u}\right)+ 
\\
      && \frac{\pi^2 c_0^4}{m_0^4}\left(\frac{8\left({M_n}^4 u^2
       \left( 7 t^3 + 31 t^2 u + 36 t u^2 + 13 u^3
         \right)  \right) }{9{M_n}^8 u} \right)
\end{eqnarray}
\begin{eqnarray}
\nonumber
T_{44} &=&  \frac{\lambda_s^2 c_0^4}{512 m_0^8}\left(\frac{ 8688
{M_n}^{16}
 - 9720 {M_n}^{14} \left( t + u \right)  + 
      9 t {\left( t - u \right) }^2 u 
       {\left( t + u \right) }^4}{9 {M_n}^8}\right) +
\\
\nonumber
&& \frac{\lambda_s^2 c_0^4}{512 m_0^8}\left(\frac{ 20 {M_n}^{10} \left(
t + u \right) 
       \left( 85 t^2 + 52 t u + 85 u^2 \right)}{9 {M_n}^8}\right)  + 
\\
\nonumber
&&\frac{\lambda_s^2 c_0^4}{512 m_0^8}\left(\frac{ {M_n}^{12}
       \left( 428 t^2 + 5048 t u + 428 u^2 \right)}{9 {M_n}^8}\right)  - 
\\
\nonumber
&&      
\frac{\lambda_s^2 c_0^4}{512 m_0^8}\left(\frac{ 12 {M_n}^2 {\left( t
+ u \right) }^3
       \left( t^4 + 8 t^3 u - 10 t^2 u^2 + 8 t u^3 + 
         u^4 \right)}{9 {M_n}^8}\right)  + 
\\
\nonumber
&&\frac{\lambda_s^2 c_0^4}{512
m_0^8}\left(\frac{4 {M_n}^8
       \left( 17 t^4 + 403 t^3 u + 24 t^2 u^2 + 
         403 t u^3 + 17 u^4 \right)}{9 {M_n}^8}\right)  - 
\\
\nonumber
&&\frac{\lambda_s^2 c_0^4}{512 m_0^8}\left(\frac{ 4 {M_n}^6 \left( t + u
\right) 
       \left( 117 t^4 + 482 t^3 u + 410 t^2 u^2 + 
         482 t u^3 + 117 u^4 \right)} {9 {M_n}^8}\right)  + 
\\
&& \frac{\lambda_s^2 c_0^4}{512 m_0^8}\left(\frac{ {M_n}^4{\left( t + u
\right)}^2
       \left( 135 t^4 + 604 t^3 u + 58 t^2 u^2 + 
         604 t u^3 + 135 u^4 \right)}{9 {M_n}^8}\right) 
\end{eqnarray}
\begin{eqnarray}
\nonumber
T_{14} &=& -\frac{\pi c_0^4 \lambda_s}{4 m_0^6}\left(\frac{  864
{M_n}^{16} + 
      3 t^4 \left( t - u \right)  u 
       {\left( t + u \right) }^2}{9 {M_n}^8 t}\right) + 
\\
\nonumber
&&\frac{\pi c_0^4 \lambda_s}{4 m_0^6}\left(\frac{ 2 {M_n}^2 t^4 \left( 3
t - u \right) 
       \left( t + u \right)  \left( t + 5 u \right)}{9 {M_n}^8
t}\right)  + 
\\
\nonumber
&& \frac{\pi c_0^4 \lambda_s}{4 m_0^6}\left(\frac{12 {M_n}^{14}
       \left( 80 t + 63 u \right)  + 
      {M_n}^{12}
       \left( -79 t^2 - 404 t u + 207 u^2 \right)}{9 {M_n}^8 t}\right)  - 
\\
\nonumber
&&      \frac{\pi c_0^4 \lambda_s}{4 m_0^6}\left(\frac{{M_n}^8 t
       \left( 39 t^3 - 49 t^2 u - 323 t u^2 - 171 u^3
         \right)}{9 {M_n}^8 t}\right)  -
\\
\nonumber
&&\frac{\pi c_0^4 \lambda_s}{4 m_0^6}\left(\frac{2 {M_n}^{10}
       \left( 153 t^3 + 593 t^2 u + 331 t u^2 - 9 u^3
         \right)}{9 {M_n}^8 t}\right)  + 
\\
\nonumber
&&\frac{\pi c_0^4 \lambda_s}{4 m_0^6}\left(\frac{2 {M_n}^6 t
       \left( 67 t^4 + 201 t^3 u + 175 t^2 u^2 + 
         37 t u^3 - 6 u^4 \right)}{9 {M_n}^8 t}\right)  - 
\\
&& \frac{\pi c_0^4 \lambda_s}{4 m_0^6}\left(\frac{{M_n}^4 t^2 
       \left( 51 t^4 + 186 t^3 u + 143 t^2 u^2 + 
         94 t u^3 + 18 u^4 \right) }{9 {M_n}^8 t} \right) 
\end{eqnarray}
\begin{eqnarray}
\nonumber
T_{24} &=& -\frac{\pi c_0^4 \lambda_s}{4 m_0^6}\left(\frac{ \left( 864
{M_n}^{16} - 
      3 t \left( t - u \right) u^4 
       {\left( t + u \right) }^2\right)}{9 {M_n}^8 u}\right) +
\\
\nonumber 
     &&\frac{\pi c_0^4 \lambda_s}{4 m_0^6}\left(\frac{ 2 {M_n}^2 \left( t
- 3 u \right)  u^4
       \left( t + u \right) \left( 5 t + u \right)}{9 {M_n}^8 u}\right)  + 
\\
\nonumber     
&&\frac{\pi c_0^4 \lambda_s}{4 m_0^6}\left(\frac{ 12 {M_n}^{14}
       \left( 63 t + 80 u \right)  + 
      {M_n}^{12} 
       \left( 207 t^2 - 404 t u - 79 u^2 \right)}{9 {M_n}^8 u}\right)  - 
\\
\nonumber
&&     
\frac{\pi c_0^4 \lambda_s}{4 m_0^6}\left(\frac{ {M_n}^8 u
       \left( -171 t^3 - 323 t^2 u - 49 t u^2 + 
         39 u^3 \right)}{9 {M_n}^8 u}\right)  -
\\ 
\nonumber
&&
     \frac{\pi c_0^4 \lambda_s}{4 m_0^6}\left(\frac{ {M_n}^{10}
       \left( -18 t^3 + 662 t^2 u + 1186 t u^2 + 
         306 u^3 \right)}{9 {M_n}^8 u}\right)  -
\\
\nonumber
&&
     \frac{\pi c_0^4 \lambda_s}{4 m_0^6}\left(\frac{ {M_n}^4 u^2 
       \left( 18 t^4 + 94 t^3 u + 143 t^2 u^2 + 
         186 t u^3 + 51 u^4 \right)}{9 {M_n}^8 u}\right)  +
\\
&& 
     \frac{\pi c_0^4 \lambda_s}{4 m_0^6}\left(\frac{ 2 {M_n}^6 u 
       \left( -6 t^4 + 37 t^3 u + 175 t^2 u^2 + 
         201 t u^3 + 67 u^4 \right)  ) }{9 {M_n}^8 u}\right) 
\end{eqnarray}
\begin{eqnarray}
\nonumber
T_{34} &=& \frac{\pi c_0^4 \lambda_s}{32 m_0^6}\left(\frac{16 \left( -588
{M_n}^{14} + 
      630 {M_n}^{12} \left( t + u \right)\right)}{9 {M_n}^8}\right)  +
\\
\nonumber
&& 
     \frac{\pi c_0^4 \lambda_s}{32 m_0^6}\left(\frac{ 3 t {\left( t - u \right) }^2 u
       {\left( t + u \right) }^3}{9 {M_n}^8}\right) - 
\\
\nonumber
&&     \frac{\pi c_0^4 \lambda_s}{32 m_0^6}\left(\frac{ {M_n}^8 \left( t +
u \right) 
       \left( 5 t^2 - 124 t u + 5 u^2 \right)}{9 {M_n}^8}\right)  - 
\\
\nonumber
&&     
\frac{\pi c_0^4 \lambda_s}{32 m_0^6}\left(\frac{ 2 {M_n}^{10} 
       \left( 43 t^2 + 244 t u + 43 u^2 \right)}{9 {M_n}^8}\right)  - 
\\
\nonumber
&&     
\frac{\pi c_0^4 \lambda_s}{32 m_0^6}\left(\frac{ {M_n}^2 {\left( t + u
\right) }^2 
       \left( 3 t^4 + 22 t^3 u - 26 t^2 u^2 + 22 t u^3 + 
         3 u^4 \right)}{9 {M_n}^8}\right)  -
\\
\nonumber
&& 
     \frac{\pi c_0^4 \lambda_s}{32 m_0^6}\left(\frac{ 4 {M_n}^6 
       \left( 8 t^4 + 61 t^3 u + 30 t^2 u^2 + 61 t u^3 + 
         8 u^4 \right)}{9 {M_n}^8}\right)  + 
\\
&& 
    \frac{\pi c_0^4 \lambda_s}{32 m_0^6}\left(\frac{ {M_n}^4 \left( t + u
\right)  
       \left( 21 t^4 + 92 t^3 u + 38 t^2 u^2 + 
         92 t u^3 + 21 u^4 \right) }{9 {M_n}^8}\right)\ .
\end{eqnarray}
Here s, t and u are the Mandelstam variables. 

\end{document}